\documentclass[12pt]{article}
\usepackage[letterpaper, portrait, margin=1in]{geometry}
\usepackage{amsfonts,amsmath,amssymb,amsthm}
\usepackage[hidelinks]{hyperref}
\hypersetup{colorlinks,citecolor=blue,linkcolor=blue,urlcolor=black}
\usepackage[hypcap]{caption}
\usepackage[utf8]{inputenc}
\usepackage[english]{babel}
\usepackage{graphicx}
\usepackage{natbib}
\usepackage{geometry}
\usepackage{booktabs}	
\usepackage{caption}
\usepackage{cases}
\usepackage{wrapfig}
\usepackage{float}
\usepackage{lscape}
\usepackage{rotating}
\usepackage{comment}
\usepackage{color}
\usepackage{multirow}
\usepackage{calc}
\usepackage[nameinlink]{cleveref}
\usepackage{setspace} 
\usepackage[shortlabels]{enumitem}
\usepackage{epstopdf}
\usepackage[final]{pdfpages}
\usepackage{csquotes}

\let\emptyset\varnothing

\usepackage[multiple]{footmisc}
\def\signed #1{{\leavevmode\unskip\nobreak\hfil\penalty50\hskip2em
  \hbox{}\nobreak\hfil(#1)%
  \parfillskip=0pt \finalhyphendemerits=0 \endgraf}}

\newsavebox\mybox

\theoremstyle{plain}
\newtheorem{thm}{Theorem}
\newtheorem{lem}{Lemma}
\newtheorem{prop}{Proposition}
\newtheorem{coro}{Corollary}

\theoremstyle{definition}

\newtheorem{ex}{Example}

\theoremstyle{remark}

\newtheorem*{claim*}{Claim}

\title{Non-Standard Choice in Matching Markets\thanks{We are deeply grateful to Tayfun S\"{o}nmez, M. Utku \"{U}nver, and M. Bumin Yenmez for guidance and support throughout this project. We thank In\'{a}cio B\'{o}, Soo Hong Chew, Umut Dur, Thayer Morrill, Josue Ortega, Bobby Pakzad-Hurson, William Phan, Bertan Turhan, Rakesh Vohra, three anonymous reviewers, and seminars participants at the 2023 Science of Decision Making Conference at HKU, Impromptu Workshop at SWUFE CCBEF, North Carolina State University, Boston College, the 2022 Easter Workshop on School Choice at Queen's University Belfast, Brown University, the 2022 Conference on Mechanism and Institution Design at the National University of Singapore, University of Texas Dallas JSOM, Southwestern University of Finance and Economics CCBEF, National Taiwan University, and Peking University HSBC Business School for helpful comments and discussions. All errors are our own. }}

\author{Gian Caspari\thanks{Department of Market Design, ZEW — Leibniz Centre for European Economic Research, Mannheim 68161, Germany. Email: gian.caspari@zew.de.}
\and Manshu Khanna\thanks{Peking University HSBC Business School, Shenzhen 518055, China. Email: manshu@phbs.pku.edu.cn.}}

\date{August 9, 2024}

\begin{document}

\maketitle

\begin{abstract}	
We explore the possibility of designing matching mechanisms that can accommodate non-standard choice behavior. We pin down the necessary and sufficient conditions on participants’ choice behavior for the existence of stable and incentive compatible mechanisms. Our results imply that well-functioning matching markets can be designed to adequately accommodate a plethora of choice behaviors, including the standard behavior consistent with preference maximization. To illustrate the significance of our results in practice, we show that a simple modification in a commonly used matching mechanism enables it to accommodate non-standard choice behavior.  
\end{abstract}



\newpage
\onehalfspacing

\section{Introduction}

This paper explores the consequences of incorporating non-standard choice behaviors, which diverge from traditional rational choice models, into the theory of stable matchings. We are motivated by significant evidence across marketing, psychology, and economics, which indicates that individuals often deviate from rational decision-making due to errors and behavioral biases.  Such behaviors include but are not limited to, choice overload, framing, and attraction effects, as well as temptation, self-control issues, and a preference for the status quo (\cite{salant2008f}, \cite{apesteguia2013choice}, \cite{gerasimou2018indecisiveness}).  The following examples demonstrate the significance of analyzing such behavior in matching markets:

\noindent \textbf{(i) Choice Complexity and Overload.}
In the context of matching kidney donors to patients, doctors often hesitate to express complete preferences due to the extensive information required to evaluate a kidney's suitability. 
In practice, a kidney $s^1$ may be chosen from a larger menu $\{s^1,s^2,s^3,\dots\}$ based on preliminary tests, while another kidney  $s^2$ would have been chosen from a smaller menu $\{s^1,s^2\}$ with more scope for a detailed examination  \citep{bade2016pareto}). Such choice behavior is non-standard, and its analysis falls beyond the scope of classical matching theory. 

\noindent \textbf{(ii) Groups as Participants.} Take the example of school admissions, where parents report a ranking over schools to a centralized authority. The preferences of parents and those of the people they consult with to make this decision may not fully align. As a result, their final decision might involve aggregating several preferences in some fashion. Naturally, such aggregation decisions need not be consistent with the maximization of a single preference relation.
Thus, participants may exhibit non-standard choice behavior even in the absence of behavioral biases and errors.

\noindent \textbf{(iii) Hiring with Attraction Effect.} Consider a hypothetical labor market choice situation where a candidate faces a decision among three job offers from different companies: $\{s^1, s^2, s^3\}$. Job $s^1$ is considered superior to job $s^2$. However, due to the attraction effect (\cite{huber1982adding}), the presence of a third option, $s^3$, which is similar yet less desirable than $s^2$, could sway the candidate's decision. As a result, the candidate might prefer $s^1$ when comparing just $\{s^1, s^2\}$, but switch their preference to $s^2$ when $s^3$ is introduced as an option, $\{s^1, s^2, s^3\}$. Again, such choices are inconsistent with maximizing a single preference relation.

This paper aims to identify the weakest conditions on choice behavior to ensure well-functioning matching markets exist and to contrast our findings with the classical setup. 
Incorporating non-standard choice behavior into matching markets raises new challenges. Notably, adapting the classical criteria of stability, efficiency, and incentive compatibility to our framework can be approached from various angles, necessitating a deliberate stance on our part. Our methodology suggests pairwise choices are inherently more meaningful than choices from larger menus, echoing the ideas behind our initial examples (i) and (iii). Furthermore, reporting the necessary choice information required to run a direct mechanism is hardly feasible in practice. 
Our analysis suggests that dynamic mechanisms can be tailored to accommodate non-standard choice behavior while minimizing the elicited preference information.

Below, we discuss our model and contributions in more detail, as well as the related literature.  \Cref{section: Model} formally introduces the model and relevant definitions. \Cref{section: Stability} states the stability result and minimal conditions on choice behavior. \Cref{section: Richness} discusses the absence of the lattice structure of the stable set and side-optimal matchings. \Cref{section: Incentives} is dedicated to studying individuals' incentives. \Cref{section: Application} presents the application. Finally, \Cref{section:discussion}, delves into a more detailed discussion of our modeled choice behavior, the proposed solution concepts, and preference reporting language. 

\subsection{Model and Contributions}

We consider a standard many-to-one admissions problem that consists of individuals and institutions. In contrast to the classical setup, where individuals have preferences over potential assignments, we equip the individuals in our model with choice functions.  
This enables us to move beyond the assumption that an individual's choices are rationalizable by an underlying preference relation.

To extend the classical conditions of stability, efficiency, and incentive compatibility to our setting where individuals have choice functions, we need to establish clear criteria for what is unacceptable and what is preferred. An individual finds an assignment \textit{unacceptable} if it is not chosen from the singleton set containing it. An individual \textit{prefers} an institution over her assignment if it chooses that institution over her assignment in a pairwise comparison. Thus, selections made from both singleton and binary menus constitute significant choices.

A matching is \textit{(pairwise) stable} if no individual is assigned an unacceptable institution, no institution is assigned an unacceptable individual, and no individual-institution pair (who are originally not matched with each other) prefer being matched with each other compared to their current assignments. A matching is considered \textit{Pareto dominated} by another if no individual prefers their assignment in the former matching over the latter, and at least one individual prefers her assignment in the latter matching over the former.  An \textit{incentive compatible mechanism} ensures that no individual can attain a preferred assignment by misreporting her choices.

Without any sophistication in choice behavior, stable matchings may not exist. We provide two necessary and sufficient conditions on individuals' choice behavior for the existence of stable matchings (\Cref{Stability_MainThm}). The first condition, \textit{weak acyclicity}, rules out the presence of strict cycles in choices for any sequence of binary menus. The second condition, \textit{acceptable-consistency}, requires that an unacceptable assignment is not chosen over an acceptable assignment when offered as a pair. We construct a deferred acceptance style algorithm that always yields a stable matching for weakly acyclic and acceptable-consistent choice functions.

To connect our model to the standard model in matching theory with (strict) preferences over potential assignments, we present a characterization of individuals' ability to strictly order alternatives in terms of \cite{plott1973path}'s \textit{path independence} (\Cref{PlottAdjusted}).  
This helps contrast the requirements of standard choice behavior with the two minimal conditions on choice behavior identified in \Cref{Stability_MainThm}. We demonstrate that weakly acyclic and acceptable-consistent choice functions cover a broader range of choice behaviors and phenomena than standard path-independent choice functions (\Cref{PlottImpliesWAAC}). Specifically, we present examples of well-documented phenomena such as attraction effects, extreme status quo biases, and framing effects—phenomena that path-independent choice functions rule out but still can be accommodated by weakly acyclic and acceptable-consistent choice functions (\Cref{subsection: StabilitySC}).

In \Cref{section: Richness}, we detail the structure of the set of stable matchings. 
We start by constructing an associated market with proxy preferences --- which mimic the choice functions of individuals as closely as possible (\Cref{ProxyMarket_Existence}). 
If choice functions are path independent the two markets yield the same set of stable matchings (\Cref{Richness_Stability2}), which is to be expected since path independent choice functions represent the standard choice behavior rationalizable by an underlying preference relation.
In contrast, if the choice functions are weakly acyclic and acceptable-consistent, even though any stable matching in the proxy admission market is stable in the original market, the converse statement does not hold (\Cref{Richness_Stability}). The idiosyncrasies associated with weakly acyclic and acceptable-consistent choice functions lead to markets where every stable matching is Pareto dominated (for individuals) by another stable matching (\Cref{Richness_Structure2}). Therefore, the lattice structure of stable matchings is absent, and there are no side-optimal matchings.

In \Cref{Incentives_MainThm}, we show that weakly acyclic and acceptable-consistent choice functions are also necessary and sufficient for stable and incentive compatible mechanisms to exist. Furthermore, these two conditions are necessary and sufficient for the existence of an (seemingly larger but surprisingly) equivalent class of individually rational, weakly non-wasteful, and incentive compatible mechanisms. 

Lastly, we present an application that shows a dynamic mechanism used in practice can be modified to accommodate non-standard choice behavior (\Cref{section: Application}). Publicly announced cut-offs demarcate each step of the procedure and circumvent the requirement of eliciting a rank-ordered list of programs from students. At each step, the cut-offs reveal the minimum score required for acceptance at each program, thus offering students a menu of programs to choose from (or switch to). \Cref{Application_Simultaneous} shows that even though the procedure offers choice menus when programs announce cut-offs simultaneously, it takes path independent choice functions for the procedure to yield stable outcomes. \Cref{Application_Sequential} shows that requiring programs to announce their cut-offs sequentially instead of simultaneously enables the mechanism to accommodate non-standard choice behavior.  
Intuitively, announcing cut-offs sequentially ensures that individuals decide between at most two alternatives at a time, so that no irrelevant options can distort their choices.

\subsection{Related Literature}

Developing mechanisms accommodating non-standard choice behavior has generated substantial research across various fields, including game theory, mechanism design, implementation theory, and industrial organization. This research has highlighted the roles of bounded rationality and choice biases in strategic decision-making (\cite{compte2019ignorance}, \cite{grubb2015consumer}). Within the domain of matching theory, two primary avenues have been explored beyond standard preferences:

\begin{enumerate}
    \item Behavioral Biases: The first avenue is driven by behavioral biases, such as rank-dependent preferences, loss aversion, overconfidence, and level-k reasoning. These biases shed light on why participants might manipulate preferences within strategy-proof matching mechanisms, contributing to instabilities and experimental anomalies (\cite{meisner2022report}, \cite{meisner2023loss}, \cite{dreyfuss2022expectations}, \cite{hakimov2022confidence},  \cite{grenet2019decentralizing}), \cite{zhang2021level}), \cite{pan2019instability}, \cite{hakimov2021experiments}). 
    In contrast, individuals in our model exhibit non-standard choice behavior without an explicit influencing mechanism in the background.

    \item Complex Preferences: The second avenue expands the scope of preferences in matching models to include a broader range of motivations. This includes analysis of matching problems with peer-dependent preferences, incomplete preferences, bounded rationality, complementarities, externalities, and indifferences in preference rankings (\cite{bade2016pareto}, \cite{kuvalekar2023matching}, \cite{erdil2017two}, \cite{hatfield2010substitutes}, \cite{sasaki1996two}, \cite{pycia2021matching}, \cite{leshno2021stable}, \cite{cox2021peer}).  \cite{kuvalekar2023matching} is the closest paper to ours, which models indecisiveness through incomplete preferences in a one-to-one matching market, analogous to allowing for empty choices from non-singleton menus in our model. Our pairwise stability notion resembles  \cite{kuvalekar2023matching}'s weak stability in that, both disallow indecisive agents to block outcomes.
    
\end{enumerate}

Our study pins down the limits of generalizing individual preferences, encoded via choice functions, to guarantee the existence of a stable matching in a many-to-one setting.
This complements the extensive work done on the generalization of institutional preferences over individuals encoded via choice correspondences (\cite{hatfield2005matching},  \cite{chambers2017choice}, \cite{hatfield2020}).

Strategic and non-strategic considerations lead to sub-optimal preference reporting, undermining the performance of real-world matching mechanisms (\cite{rees2018suboptimal}, \cite{rees2023behavioral}). Explanations ranging from mistakes to preference reversals have been proposed to explain the reporting behavior that strategic considerations cannot explain (\cite{narita2018match}, \cite{shorrer2023dominated}). We consider this as suggestive evidence for non-standard choice, which provides another alternative explanation for the observed behavior. Our study corroborates the argument that dynamic mechanisms offer a promising avenue to address behavioral considerations (\cite{bo2020iterative}, \cite{Klijnetal2019}, \cite{bo2020pick}, \cite{mackenzie2020menu}).

\section{Model}\label{section: Model}
We introduce a many-to-one matching model that consists of institutions and individuals. We deviate from the standard modeling approach by equipping the individuals with choice functions instead of preference relations, allowing for more general choice behavior. 

\medskip
\noindent An \textbf{admissions problem} $\gamma\in \Gamma$ is a five-tuple $\langle I, S, q, C, \pi \rangle$ that consists of: 
\begin{enumerate}[(i)]
\item a non-empty finite set of individuals $I$, 
\item a non-empty finite set of institutions $S$, 
\item a list of capacities of institutions $q=(q_s)_{s\in S}$,
\item a list of priority orders of institutions $\pi=(\pi_s)_{s\in S}$ over $I\cup \{\emptyset\}$, and
\item a list of choice functions of individuals $C=(C_i)_{i\in I}$ over $2^S$.
\end{enumerate}

Each institution $s$ has a capacity of $q_s$ seats representing the maximum number of individuals it can accept. Priority order $\pi_s$ represents the way institution $s$ ranks individuals. Formally, a \textbf{priority order} $\pi_s$ is a strict simple order (transitive, asymmetric, and complete) over $I\cup \{\emptyset\}$. Let $\Pi$ denote the set of all possible lists of priority orders. From an institutional viewpoint, we assume that there are no complementarities between individuals, so the priority order $\pi_s$ and capacity $q_s$ of an institution $s$ translate into a (partial order) preference over sets of individuals in a straightforward way.\footnote{In a nutshell, an institution chooses the $q_s$ highest priority individuals from any set of acceptable individuals. Formally, let $\succ_s$ be a partial order over $2^I$. We assume that $\succ_s$ is responsive (see \Cref{groupstability} in \Cref{section: AppendixProofs}).}

Each individual $i$ is equipped with a choice function $C_i$ that represents her choice from any menu of institutions.  Formally, a (unit demand) \textbf{choice function} $C_i$ is a mapping $C_i: 2^S \to 2^S$ such that for every  $S'\in 2^S$ we have $C_i(S') \subseteq S'$ and $|C_i(S')|\leq 1$. 

Let us define a few basic terms. An institution $s$ is \textbf{acceptable to individual} $i$ if $C_i(\{s\})=\{s\}$ and unacceptable if $C_i(\{s\})=\emptyset$. Similarly, an individual $i$ is \textbf{acceptable to institution} $s$ if $i \mathrel{\pi}_s \emptyset$ and unacceptable if $ \emptyset \mathrel{\pi}_s I$.
A matching is an assignment of individuals to institutions, such that each individual is assigned at most one institution, and no institution is assigned more individuals than its capacity. Formally, a (feasible) \textbf{matching} is a correspondence $\mu: I \cup S \mapsto I \cup S \cup \{\emptyset\}$ that satisfies:
\begin{enumerate}[(i)]
\item $\mu(i) \subseteq S$ such that  $|\mu(i) | \leq 1$ for all $i \in I$,
\item $\mu(s) \subseteq I$ such that  $|\mu(s) | \leq q_s$ for all $s \in S$, and
\item $i \in \mu(s)$ if and only if $s \in \mu(i)$ for all $i \in I$ and $s \in S$.
\end{enumerate}
Let $\mathcal{M}$ denote the set of all (feasible) matchings.  A matching is (pairwise) stable if no individual (or institution) is assigned an unacceptable institution (or individual), and there is no individual-institution pair (who are originally not matched with each other) prefer being matched with each other possibly instead of some of their current assignments.
Formally, a matching $\mu$  is \textbf{(pairwise) stable} if
 
\begin{enumerate}[(i)]
\item it is \textbf{individually rational}, that is, there is no individual $i$ such that $C_i(\mu(i))=\emptyset$ and no institution $s$ such that $\emptyset  \mathrel{\pi_s} i$ for some $i\in \mu(s)$, and 
\item there is no \textbf{blocking pair}, that is, there is no pair $(i,s)\in I\times S$ such that
 \begin{enumerate}[(a)]
 \item  $\mu(i) \neq \{s\}$, 
 \item $C_i(\mu(i)\cup  \{s\})=\{s\}$, and
 \item 
 	\begin{enumerate}[(1)]
 	\item either  $i  \mathrel{\pi_s} i'$ for some $i'\in \mu(s)$, or 
 	\item $|\mu(s)|<q_s$ and $i  \mathrel{\pi_s} \emptyset$.
	\end{enumerate}
 \end{enumerate}

\end{enumerate}

Analogous to the classical setup, pairwise stability is equivalent to the seemingly more involved notion of \textit{group stability} (see \Cref{groupstability} in \Cref{section: AppendixProofs}), which rules out coalitions consisting of multiple individuals and institutions that prefer being matched with each other compared to their current assignments. 

Finally, a \textbf{mechanism} is a function $\psi:  \Gamma  \rightarrow \mathcal{M}$ that assigns a matching $\psi[\gamma] \in \mathcal{M}$ to each admission problem $\gamma\in \Gamma$. A mechanism  $\psi$ is \textbf{stable} if $\psi[\gamma] $ is stable for any admission problem $\gamma\in \Gamma$.

\section{Stable Matchings}
\label{section: Stability}

\subsection{Stable Matchings under Non-Standard Choice}
\label{subsection: StabilityNSC}

Without any sophistication in choice behavior, stable matchings may not exist. For instance, a choice function that selects unacceptable alternatives over acceptable ones would violate individual rationality. This subsection describes the weakest conditions on choice behavior that guarantee the existence of stable matchings.

The first condition, weak acyclicity, rules out the possibility that an individual can always find another institution to block with, regardless of the institution assigned to it. Formally, choice function $C_i$ is \textbf{weakly acyclic (over acceptable institutions)} if for all positive integer $t \geq 3$ and $t$ distinct and acceptable institutions $s^1, s^2, \dots , s^t\in S$,\footnote{Weak acyclicity resembles \textit{Strong Axiom of Revealed Preference (SARP)}, however there are important distinctions. Unlike SARP, weak acylicty does not imply \textit{independence of irrelevance alternatives (IIA)} because it does not restrict choices from menus of size larger than two. This is shown in \Cref{Example: AttractionEffect}. For definitions and an excellent exposition of SARP and IIA, see \cite{bossert2010consistency}.}
\[C_i(\{s^1,s^2\}) = \{s^1\}, \dots, C_i(\{s^{t-1},s^{t}\}) = \{s^{t-1}\}  \text{ implies } C_i(\{s^{1},s^{t}\}) \neq \{s^{t}\}.\]

The second condition, acceptable-consistency, ensures that an individual does not choose an unacceptable institution over an acceptable one in pairwise comparisons. Formally, choice function $C_i$ is \textbf{acceptable-consistent} if for all distinct institutions $s,s'\in S$,
\[
C_i(\{s\})=\{s\}  \text{ and } C_i(\{s'\})=\emptyset  \text{ implies } C_i(\{s,s'\})\neq \{s'\}.\]

The following result shows that, for an individual with a weakly acyclic and acceptable-consistent choice function, any set of institutions with at least one acceptable institution contains at least one institution such that no other institution in the set is chosen over in pairwise comparisons. Formally, let the set of \textbf{C-maximal institutions} for individual $i$ in subset $S'\subseteq S$ be denoted by $U_i(S')\equiv \big\{s\in S': \text{ not exists } s'\in S'\setminus \{s\} \text{ such that } C_i(\{s,s'\})=\{s'\} \big\}$. The following lemma holds.

\begin{lem}
\label{Stability_Lemma}
For a weakly acyclic and acceptable-consistent choice function, $C_i$ and a subset of institutions $S'\subseteq S$ containing at least one acceptable institution,  the set of C-maximal institutions $U_i(S')$ is non-empty. 
\end{lem}

An analogous but much stronger version of this result trivially holds for the case of strict preferences. That is, every set of institutions, with at least one acceptable institution, must contain an institution preferred to any other institution in the set. Our next result shows that the weaker version presented in \Cref{Stability_Lemma} is enough to construct a deferred acceptance style mechanism that always leads to a stable outcome. Moreover, if either weak acyclicity or acceptable-consistency are violated, the existence of a stable matching is no longer guaranteed. 

\begin{thm}
\label{Stability_MainThm}
Fix $I, S, q, C$. There exists a stable matching for every priority order profile $\pi\in \Pi$ if and only if the choice functions are weakly acyclic and acceptable-consistent. 
\end{thm}

The proof of \Cref{Stability_MainThm}  in \Cref{section: AppendixProofs} describes a version of the deferred acceptance algorithm that always yields a stable matching for weakly acyclic and acceptable-consistent choice functions. At each step of the original (individual proposing) deferred acceptance \citep{gale1962college}, every individual not currently tentatively accepted by an institution proposes to its most preferred institution among the remaining institutions (that is, the ones that have not rejected her at a previous step). In contrast, a single top choice might not exist in our setting with non-standard choices. Instead, our algorithm ensures that any individual proposes to an institution such that she would not choose any remaining institution over it in a pairwise comparison.  Given that the choice functions are weakly acyclic and acceptable-consistent by \Cref{Stability_Lemma}, such an institution exists in any non-empty set of institutions containing at least one acceptable institution.

This result highlights that matching markets can be designed to accommodate many choice behaviors that are not allowed under the classical setup consisting of individuals with preference relations. However, the exact connection between preference relations and weak acyclic and acceptable-consistent choice functions remains to be established. This is the point of the next subsection.

\subsection{Standard Assumptions on Choice Behavior}\label{subsection: StabilitySC}

The two identified conditions are new to the literature on stable matching theory. The standard assumption in the literature is that individuals can rank the institutions (together with the option of remaining unassigned) in a single order. We next show that the ability to rank institutions corresponds to a well-known condition on choice functions in our setup called path independence. 

Let us first define path independence formally. A choice function is \textbf{path independent} if for all $S',S''\subseteq S$ we have
\begin{center}
 $C_i(S'\cup S'')=C_i(C_i(S')\cup S'')$.
\end{center}

Path independence requires that if a set is segmented arbitrarily, choice from the menu consisting of only the chosen assignments from each segment, must be the same as the choice made from the unsegmented set. We next show that a path independent choice function reflects choice behavior that a strict order over institutions can rationalize. That is, given any path independent choice function, there exists a unique strict order over (acceptable) institutions such that the institution chosen from each menu of institutions is simply the best in that menu with respect to the strict order. This benchmark result allows for a direct comparison of the perturbed setup consisting of the two conditions identified in \Cref{Stability_MainThm} with the classical setup of stable matching theory.

Let us proceed to make the idea of choice behavior consistent with the ability to rank institutions formal. Let $R_i$ be a binary relation over $S \cup \{\emptyset\}$. 
A binary relation $R_i$ over $S \cup \{\emptyset \}$ is  \textbf{strongly complete over acceptable alternatives} if
	\begin{enumerate}[(i)]
		\item for all $s\in S$ either $s \mathrel{R_i} \emptyset$ or $\emptyset \mathrel{R_i} s$, and
		\item for all $s', s''\in \{s\in S: s \mathrel{R_i} \emptyset\}$ either  $s' \mathrel{R_i} s''$ or $s'' \mathrel{R_i} s'$.
	\end{enumerate}
A binary relation is \textbf{anti-symmetric over acceptable choices} if for all $s',s''\in \{s\in S: s \mathrel{R_i} \emptyset\}$, $s' \mathrel{R_i}s''$ and $s'' \mathrel{R_i} s'$ implies $s'=s''$.
Finally, a binary relation is \textbf{transitive over acceptable choices} if for all $s',s'',s'''\in \{s\in S: s \mathrel{R_i} \emptyset\}$ we have that $s' \mathrel{R_i} s''$ and $s'' \mathrel{R_i} s'''$ implies $s'\mathrel{R_i} s'''$.
We say $R_i$ is a \textbf{simple order over acceptable choices} in  $S \cup \{\emptyset\}$ if it is strongly complete, anti-symmetric, and  transitive over acceptable choices. 
A choice function $C_i$ can be \textbf{rationalized by a simple order over acceptable choices} $R_i$  if and only if for all subsets $S'\subseteq S$, we have 
\begin{enumerate}[(i)]
    \item $C_i(S')=\emptyset$ if $\emptyset \mathrel{R_i} s$ for all $s\in S'$, and
    \item $C_i(S')=\big\{s'\in S': s' \mathrel{R_i} s'' \text{ for all } s''\in  \{s\in S: s \mathrel{R_i} \emptyset\}  \big\} $  otherwise.
\end{enumerate}

Notice that our choice functions allow for the possibility of choosing nothing (the empty set). This possibility is not present in \cite{plott1973path}'s original analysis of path independent choice functions where individuals' choice from a non-empty menu is not allowed to be empty. The possibility of empty choices from non-empty menus requires some additional careful considerations that lead to the following result.

\begin{thm}
\label{PlottAdjusted}
A (unit demand) choice function $C_i$ can be rationalized by a simple order over acceptable choices if and only if it is path independent. 
\end{thm}

 In \Cref{PlottImpliesWAAC} we  show that path independence is a stronger requirement on choice sophistication than weak acyclicity and acceptable-consistency combined.

\begin{prop}\label{PlottImpliesWAAC}
If a choice function is path independent, then it is weakly acyclic and acceptable-consistent. The converse statement may not hold. 
\end{prop}

\begin{coro} \label{StabilityCoroPI}
Fix $I, S, q, C$. There exists a stable matching for every priority order profile $\pi\in \Pi$ if the choice functions are path independent.
\end{coro}

 Let us revisit the attraction effect (\cite{huber1982adding}) example (iii) from the introduction to illustrate choice behavior that satisfies weak acyclicity and acceptable-consistency but not path independence.

 \begin{ex}[Hiring with Attraction Effect] \label{Example: AttractionEffect}
Consider an individual choosing among three institutions $\{s^1, s^2, s^3\}$, each having a vacant position. Institution $s^1$ is better than institution $s^2$. The individual's choice may be influenced by the availability of a third institution $s^3$, which is similar but inferior to $s^2$. 
That is, consider an admissions problem $\gamma=\langle I, S, q, C, \pi \rangle$ with

\begin{enumerate}[(i)]
    \item $I=\{i\}$, 
    \item $S=\{s^1, s^2, s^3\}$,
    \item $q_{s}=1$ for all $s \in S$,  
     \item $C_i(\{s\})=\{s\}$ for all $s \in S$, \\
        	$C_i(\{s^1,s^2\})=\{s^1\}$,\\
 	$C_i(\{s^1,s^3\})=\{s^1\}$, \\
 	$C_i(\{s^2,s^3\})=\{s^2\}$, \\
 	 $C_i(\{s^1,s^2,s^3\})=\{s^2\}$, and

    \item $i  \mathrel{\pi}_s \emptyset$ for all $s \in S$.

\end{enumerate}

Four simple observations follow. First, $C_i$ is not path independent because $C_i(\{s^1,s^2,s^3\})=\{s^2\} \neq C_i\big(C_i(\{s^1,s^3\}) \cup \{s^2\}\big)=\{s^1\}$, that is, the choices cannot be rationalized by a single preference relation. Second, $C_i$ is weakly acyclic and acceptable-consistent. Third, there exists a stable matching for this market, in particular,  $\mu(i)=\{s^1\}$.\footnote{We only state the minimal necessary information to define a matching in examples. In this case, notice that by the definition of matching, $\mu(i)=\{s^1\}$ implies that $\mu(s^1)=\{i\}$ and $\mu(s^2)=\mu(s^3)=\emptyset$.}
Finally, a stable matching also exists for all other priority profiles by  \Cref{Stability_MainThm} since $C_i$ is weakly acyclic and acceptable-consistent.

\end{ex}

Similar examples can be constructed for other choice behaviors that exhibit context effects, such as:

\begin{enumerate}
    \item \textbf{Extreme Status Quo.} (See \cite{apesteguia2013choice}, \cite{masatlioglu2005rational}.) A default (or status quo) institution $s^4$ biases the individual against choosing institutions in $A_{s^4} \subset S$. When her feasible choices $B$ include $s^4$, she ignores everything in $A_{s^4} \cap B$ and maximizes her total preference $P$ on $B \backslash A_{s^4}$.

    Consider $P: s^1 \succ s^2   \succ s^3 \succ \emptyset \succ s^4$ and $A_{s^4} = \{s^1\}$.

     $C_i(\{s\})=\{s\}$ for all $s \in \{s^1,s^2,s^3\}$, $C_i(\{s^4\})=\emptyset$, \\
       $C_i(\{s^1,s^2\})=\{s^1\}$, 	$C_i(\{s^2,s^3\})=\{s^2\}$,    	$C_i(\{s^3,s^4\})=\{s^3\}$,  	 $C_i(\{s^1,s^4\})=\emptyset$, and\\ $C_i(\{s^1,s^2, s^4\})=\{s^2\}.$

Notice that $C_i$ is weakly acyclic and acceptable-consistent, but not path independent since $C_i(\{s^1,s^2, s^4\})\not= C_i(\{s^1,s^2\} \cup C_i(\{s^4\}) )$.

\item \textbf{Framing.} (See \cite{salant2018contracts}.) Because of framing, the individual chooses from the menu of institutions $B$ according to a total preference $P_f$ that differs from her ``true" total preference P. After the framing effect dissipates, she rejects the institution from $B$ unless it is preferred to her outside option $\emptyset$ according to $P$.

    Consider $P_f: s^4 \succ_f s^1 \succ_f s^2   \succ_f s^3 \succ_f \emptyset$, and \\ $P: s^1 \succ s^2   \succ s^3 \succ \emptyset \succ s^4$.

         $C_i(\{s\})=\{s\}$ for all $s \in \{s^1,s^2,s^3\}$, $C_i(\{s^4\})=\emptyset$, \\
       $C_i(\{s^1,s^2\})=\{s^1\}$, 	$C_i(\{s^2,s^3\})=\{s^2\}$,    	$C_i(\{s^3,s^4\})=\emptyset$,  	 $C_i(\{s^1,s^4\})=\emptyset$, and\\ $C_i(\{s^1,s^2, s^3, s^4\})=\emptyset.$

       Notice that $C_i$ is weakly acyclic and acceptable-consistent, but not path independent since $C_i(\{s^1,s^2, s^3, s^4\})\not= C_i(\{s^1,s^2\} \cup C_i(\{s^3,s^4\}) )$.

\end{enumerate}

Although weak acyclicity and acceptable-consistency allow for more general choice behavior than path independence,
not all known behavior biases can be accommodated. For example, consider \cite{manzini2007sequentially}'s choice behavior which can lead to cycles in pairwise choices and thus violates weak-acyclicity.

\section{Richness}  \label{section: Richness}

\subsection{Stable Matchings}

We show that the set of stable matchings under weakly acyclic and acceptable-consistent choice functions is richer than the set of stable matchings in the classical setup. 
We start by constructing an associated proxy admission problem that differs only in that individuals have preference relations as opposed to choice functions, with the preferences of individuals mimicking their choices from binary menus.
Let us proceed to the definitions.  Let $P_i$ be a binary relation over $S \cup \{\emptyset\}$. 
A binary relation $P_i$ over $S\cup \{\emptyset \}$ is \textbf{complete over acceptable alternatives} if
	\begin{enumerate}[(i)]
		\item for all $s\in S$ either $s \mathrel{P_i} \emptyset$ or $\emptyset \mathrel{P_i} s$, and
		\item for all $s', s''\in \{s\in S: s \mathrel{P_i} \emptyset\}$ such that $s' \not=s''$ either  $s' \mathrel{P_i} s''$ or $s'' \mathrel{P_i} s'$.
	\end{enumerate}
A binary relation is \textbf{asymmetric over acceptable choices} if for all $s',s''\in \{s\in S: s \mathrel{P_i} \emptyset\}$, $s' \mathrel{P_i}s''$ implies $\neg(s'' \mathrel{P_i} s')$.
We say $P_i$ is a \textbf{strict simple order over acceptable choices} in  $S \cup \{\emptyset\}$ if it is complete, asymmetric, and transitive over acceptable choices. 
We slightly abuse notation by allowing the strict simple order to compare elements with singleton sets. Specifically, we write \( s \mathrel{P_i} \mu(i) \) instead of \( s \mathrel{P_i} s' \) where \(\mu(i) = \{s'\}\).

Let $\gamma_P=\langle I, S, q, P, \pi \rangle$ be a \textbf{proxy admissions problem}  for admission problem $\gamma=\langle I, S, q, C, \pi \rangle$, where $P$ is a profile of strict simple orders over acceptable choices satisfying the following conditions: 
\begin{enumerate}[(i)]
    \item If $C_i(\{s,s'\})=\{s\}$ then $s \mathrel{P_i} s'$,

    \item If $C_i(\{s\})=\{s\}$ then $s \mathrel{P_i} \emptyset$, and
    \item If $C_i(\{s\})=\emptyset$ then $\emptyset \mathrel{P_i} s$.
\end{enumerate}

Let $\Gamma_{\gamma}$ denote the set of all proxy admissions problems for admission problem $\gamma$. In order to make sure that an associated market constructed in such a way always exists, we need to check that $\Gamma_{\gamma}$ is non-empty. For admissions problems with weakly acyclic and acceptable-consistent choice functions, the answer is affirmative.

\begin{lem}\label{ProxyMarket_Existence}
For an admissions problem $\gamma$ with choice functions that are weakly acyclic and acceptable-consistent, the set of proxy admissions problems $\Gamma_{\gamma}$ is non-empty.
\end{lem}

The following result shows that weak acyclicity and acceptable-consistency are not only weaker than path independence but that they also yield a larger set of stable matchings. Let us first define stability for proxy problems. 
A matching $\mu$ is \textbf{stable for proxy admissions problem} $\gamma_P$ if
\begin{enumerate}[(1)]
\item it is \textbf{individually rational}, that is, there is no individual $i$ such that $\emptyset \mathrel{P_i} \mu(i)$ and no institution $s$ such that $\emptyset \mathrel{\pi_s} i$ for some $i\in \mu(s)$, and 
\item there is no \textbf{blocking pair}, that is, there is no pair $(i,s)\in I\times S$ such that
 \begin{enumerate}
 \item $s \mathrel{P_i} \mu(i)$, and
 \item 
 	\begin{enumerate}[(i)]
 	\item either  $i\mathrel{\pi_s} i'$ for some $i'\in \mu(s)$, or 
 	\item $|\mu(s)|<q_s$ and $i\mathrel{\pi_s} \emptyset$.
	\end{enumerate}
 \end{enumerate}

\end{enumerate}

With weakly acyclic and acceptable-consistent choice functions, there are stable matchings of some original admissions problems that are not stable in any associated proxy admissions problem. 

\begin{prop}\label{Richness_Stability}
Fix an admissions problem $\gamma$ with choice functions that are weakly acyclic and acceptable-consistent. 
\begin{enumerate}
    \item If a matching is stable for some associated proxy admissions problem, then it is also stable for the admissions problem. 
    \item The converse statement may not hold.
\end{enumerate}
\end{prop}

The following example shows that the converse does not hold:

\begin{ex}\label{Richness_Example1}
Consider an admissions problem $\gamma=\langle I, S, q, C, \pi \rangle$ with
\begin{enumerate}[(i)]
    \item $I=\{i\}$,
    \item $S=\{s^1, s^2, s^3\}$,
    \item $q_{s}=1$ for all $s \in S$, 
     \item $C_i(\{s\})=\{s\}$ for all $s \in S$,\\
    $C_i(\{s^1,s^2\}) =\{ s^1\}$,\\
    $C_i(\{s^2,s^3\}) = \{s^2\}$,\\
    $C_i(\{s^1,s^3\}) = \emptyset$,

    \item $  i \mathrel{\pi}_{s^1} \emptyset$,\\
         $  \emptyset \mathrel{\pi}_{s^2} i$, and\\
        $  i \mathrel{\pi}_{s^3} \emptyset$.
\end{enumerate}

A few simple observations follow. First, under the original admissions problem both the matchings $\mu(i)=\{s^1\}$ and  $\nu(i)=\{s^3\}$ are stable. Second, notice that there is a unique associated proxy admissions problem for this problem. The only preference consistent with these choices is $P_{i}$ such that $s^1\mathrel{P_{i}}s^2\mathrel{P_{i}}s^3$. Finally, for the associated proxy problem  $\mu(i)=\{s^1\}$ is the only stable matching. $\nu(i)=\{s^3\}$ is not stable as it is blocked by $s^1$ and $i$ in the proxy problem.
\end{ex}

This is not the case for path independent choice behavior. Our next result shows that, for admissions problems with path independent choice functions, the set of stable matchings for any admission problem is identical to the set of stable matchings for its associated proxy admission problems, affirming that the change in the set of stable matchings in \Cref{Richness_Stability} can be solely attributed to weakening assumptions on choice behavior.

\begin{prop}\label{Richness_Stability2}
Fix an admissions problem $\gamma$ with choice functions that are path independent.
\begin{enumerate}
    \item If a matching is stable for some associated proxy admissions problem, then it is also stable for the admissions problem. 
    \item The converse statement holds.
\end{enumerate}
\end{prop}

\subsection{Constrained Efficient Stable Matching}

In the classical setup consisting of individuals with preference relations without indifferences there exists a unique \textit{individual-optimal stable matching} (\cite{gale1962college}) that every individual (weakly) prefers to any other stable matching. For matching markets consisting of individuals allowed to have indifferences, a weakening of the notion of individual-optimality, called \textit{constrained efficiency}, is seen in \cite{erdil2008s}. A constrained efficient stable matching corresponds to a matching not Pareto dominated by any other stable matching. We next define this property formally and then show that sufficient indecisiveness on the side of individuals, even a constrained efficient matching does not exist (\Cref{Richness_Structure2}).

A stable matching $\mu$ is \textbf{constrained efficient (for individuals)} if it is not Pareto dominated by any other stable matching $\mu'$. That is, $\mu'$ \textbf{Pareto dominates (for individuals)} $\mu$ if $C_i(\mu'(i)\cup \mu(i))\not=\mu(i)$ for every $i \in I$ with $\mu'(i)\neq \mu(i)$, and $C_j(\mu'(j)\cup \mu(j))=\mu'(j)$ for some $j \in I$ with $\mu'(j)\neq \mu(j)$.

\begin{prop}\label{Richness_Structure2}
There exists an admissions problem $\gamma \in \Gamma$ with choice functions that are weakly acyclic and acceptable-consistent, that does not have a constrained efficient stable matching.
\end{prop}

 The following example proves the result:

\begin{ex}
\label{Example:ConstrainedCycle}
Consider an admissions problem $\gamma=\langle I, S, q, C, \pi \rangle$ where
\begin{enumerate}[(i)]
    \item $I=\{i^1, i^2,i^3\}$, 
    \item $S=\{s^1, s^2, s^3\}$, 
    \item $q_{s}=1$ for all $s \in S$,
     \item $C_i(\{s\})=\{s\}$ for all $i \in I$ and $s \in S$. Moreover,

\begin{table}[H]

\begin{tabular}{cc|c|c|c|}
  & \multicolumn{1}{c}{} & \multicolumn{3}{c}{Choices} \\
  & \multicolumn{1}{c}{} & \multicolumn{1}{c}{$C_{i^1}$}  & \multicolumn{1}{c}{$C_{i^2}$}  & \multicolumn{1}{c}{$C_{i^3}$} \\\cline{3-5}
            & $\{s^1, s^2\}$ & $ \emptyset $ & $ s^1 $ & $ s^1$ \\ \cline{3-5}
$\quad \quad $ Choice Menu  & $\{s^2, s^3\}$ & $\emptyset$ & $ s^2 $ & $ s^2$ \\\cline{3-5}
            & $\{s^1, s^3\}$ & $ \emptyset $ & $ \emptyset $ & $ \emptyset$ \\\cline{3-5}
\end{tabular}
\end{table}

    \item $ i^1  \mathrel{\pi}_{s} i^2 \mathrel{\pi}_{s} i^3$  for all $s \in S$.

\end{enumerate}
   
Note that $i^1$ would never block a matching where she is assigned some institution. Similarly, $i^3$ having the lowest priority, can never block any other individual's assignment. Moreover, $i^2$ can only block assignments of $i^3$. It follows that there are exactly four stable matchings for problem $\gamma$. 

\begin{enumerate}[(i)]
    \item $\mu(i^1)=\{s^2\}$, $\mu(i^2)=\{s^1\}$ and $\mu(i^3)=\{s^3\}$;
      \item $\kappa(i^1)=\{s^1\}$, $\kappa(i^2)=\{s^2\}$ and $\kappa(i^3)=\{s^3\}$; 
    \item  $\eta(i^1)=\{s^2\}$, $\eta(i^2)=\{s^3\}$ and $\eta(i^3)=\{s^1\}$; and

    \item $\nu(i^1)=\{s^3\}$, $\nu(i^2)=\{s^1\}$ and $\nu(i^3)=\{s^2\}$.
    
\end{enumerate}

Its easy to check that, $\nu$ Pareto dominates $\mu$,  $\mu$ Pareto dominates $\kappa$, $\kappa$ Pareto dominates $\eta$, and $\eta$ Pareto dominates $\nu$. Also, $\nu$ Pareto dominates $\kappa$, $\kappa$ Pareto dominates $\eta$, and $\eta$ Pareto dominates $\nu$.  Since every stable matching has a Pareto improvement which is still stable, a constrained efficient stable matching does not exist in this case. 
\end{ex}

\section{Incentives} \label{section: Incentives}

Turning to analyzing individuals' incentives, we show that weakly acyclic and acceptable-consistent choice functions are necessary and sufficient for a mechanism that is both stable and incentive compatible.

A mechanism is incentive compatible (for individuals) if misreporting choices does not lead to a better assignment.
Formally, a mechanism $\psi$ is \textbf{incentive compatible (for individuals)} if for any admissions problem $\gamma=\langle I, S, q, C, \pi \rangle$ there does not exist $\hat{\gamma}=\langle I, S, q, (\hat{C}_i, C_{-i}), \pi \rangle$  such that
\begin{align*}
\psi[\gamma](i) \neq \psi[\hat{\gamma}](i)  \quad \text{and} \quad   C_i(\psi[\hat{\gamma}](i)\cup \psi[\gamma](i))= \psi[\hat{\gamma}](i).
\end{align*}

In cases where stability is not a concern, one might still be interested in the existence of incentive compatible mechanisms. We show that weakly acyclic and acceptable-consistent choice functions are sufficient for the existence of incentive compatible mechanisms, and under two additional mild conditions are also necessary.

First, we need individual rationality to ensure the mechanism assigns an acceptable institution to every individual. This rules out trivial incentive compatible mechanisms that assign every individual the same institution regardless of choices reported. Formally, a matching $\mu$ is {individually rational} if there is no individual $i$ such that $C_i(\mu(i))=\emptyset$ and no institution $s$ such that $\emptyset \pi_s i$ for some $i\in \mu(s)$. A mechanism $\psi$ is \textbf{individually rational} if $\psi[\gamma] $ is individually rational for any admission problem $\gamma\in \Gamma$.

Second, we need to ensure that no unassigned individual prefers an institution with one or more empty slots and where she is acceptable. This rules out trivial incentive compatible and individual rational mechanisms that leave all individuals unassigned. Formally, a matching $\mu$ is {weakly non-wasteful} if there exists no individual $i$ with $\mu(i)=\emptyset$ and $s$ with $|\mu(s)|<q_s$ such that $C_i(\{s\})=\{s\}$ and $s \pi_s \emptyset$. A mechanism $\psi$ is \textbf{weakly non-wasteful} if $\psi[\gamma] $ is weakly non-wasteful for any admission problem $\gamma\in \Gamma$. It is easy to see that a stable matching is always weakly non-wasteful. 
Let us formally state the result.

\begin{thm} 
\label{Incentives_MainThm}
Fix $I, S, q, C$. Then the following statements are equivalent:  
\begin{enumerate}
    \item The choice functions are weakly acyclic and acceptable-consistent.
    \item There exists a stable and incentive compatible mechanism for every priority order profile $\pi \in \Pi$.
    \item There exists an individually rational, weakly non-wasteful, and incentive compatible mechanism for every priority order profile $\pi \in \Pi$.

\end{enumerate}
\end{thm}

\section{Application}\label{section: Application}

This section shows that dynamic mechanisms can be tailored to adequately accommodate non-standard choice behavior by reducing the size of the encountered choice sets. We examine a dynamic college admissions process inspired by the University of Delhi, mirroring the steps of the Deferred Acceptance algorithm. This process starts with high-school graduates applying to colleges and reporting national exam scores, which the university uses to set cut-off marks for program admissions. Applicants must select programs whose scores exceed these cut-offs and secure their spot by submitting necessary documents within a designated time frame. If subsequent rounds offer lower cut-offs, allowing for more favorable program options, students can permanently switch programs, forfeiting their previous choice. This iterative process, requiring applicants to reassess their options based on new cut-off lists constantly, highlights the need for sophisticated decision-making to achieve stable outcomes, emphasizing the importance of path-independent choice behavior (\Cref{Application_Simultaneous}). 

To integrate the University of Delhi's admissions procedure into our model, we propose a schematic algorithm for computing the matching, with a notable modification: In our model, the programs prioritize applicants based on merit scores. 
Let the \textbf{merit score} of individual $i$ at institution $s$ be defined as 

\begin{equation*}
    m_s(i)= 
\begin{cases}
     | \{i'\in I: i \pi_s i'\}|+1, & \text{if } i \mathrel{\pi_s} \emptyset\\
    0,              & \text{otherwise}
\end{cases}
\end{equation*}

We now define the mechanism formally. 

\subsubsection*{Admissions using Simultaneous Cut-offs Algorithm }

\begin{itemize}
    \item \textbf{Step 0:} Let $\psi[\gamma]^0(i) \equiv \emptyset$, $\psi[\gamma]^{0}(s)  \equiv \emptyset$  and $c^0_s \equiv n+1$.

\item \textbf{Step t (t $\geq$ 1) with Simultaneous Cut-offs:} If each institution $s$ has filled its capacity or has announced the lowest cut-off possible ($c_s=1$),  the algorithm stops and returns $\psi[\gamma]^{t-1}$. That is, tentative assignments become permanent.

\textbf{Each} institution $s$ with $c^{t-1}_{s}>1$ that has not filled its capacity, announces the lowest cut-off $c^t_s \in \mathbb{N}$ such that  $|\{i\in I: c^t_s \leq m_s(i) < c^{t-1}_s\}| \leq q_s - | \psi[\gamma]^{t-1}(s)| $.  Other institutions announce $c^{t}_{s'} = c^{t-1}_{s'}$.

Let $B^t_i \equiv \{s\in S: c^t_s \leq m_s(i) < c^{t-1}_s \}\cup \psi[\gamma]^{t-1}(i) $ denote the set of institutions that individual $i$ can choose from. Individual $i$'s tentative assignment is updated as follows:

\begin{equation*}
    \psi[\gamma]^{t}(i)= 
\begin{cases}
    C_i(B^t_i) ,& \text{if } C_i(B^t_i )\neq \emptyset\\
    \psi[\gamma]^{t-1}(i),           & \text{otherwise}
\end{cases}
\end{equation*}

Institution $s$'s assignment is $\psi[\gamma]^{t}(s)=\{i \in I: \psi[\gamma]^t(i)=\{s\}\}$. 
\end{itemize}

The algorithm terminates in a finite number of steps. Let $\psi^{sim}$ be the mechanism based on the simultaneous cut-offs algorithm. The following result formalizes the requirements of this mechanism in terms of the choice sophistication of the applicants.

\begin{prop}\label{Application_Simultaneous}
The simultaneous cut-offs mechanism $\psi^{sim}$ leads to a stable outcome $\psi^{sim}[\gamma]$ for every admissions problem $\gamma\in \Gamma$ if and only if the choice functions are path independent.
\end{prop}

This result highlights that not all dynamic mechanisms handle choice complexity proficiently.
We simplify the decision process by altering the mechanism to release program cut-offs sequentially rather than simultaneously. This adjusted mechanism presents individuals with two programs at each step, so applicants can easily pick the more preferred option in a binary comparison. 
We now define this mechanism formally.

\subsubsection*{Admissions with Sequential Cut-offs Algorithm}

\begin{itemize}
    \item \textbf{Step 0:} Let $\psi[\gamma]^0(i) \equiv \emptyset$, $\psi[\gamma]^{0}(s)  \equiv \emptyset$  and $c^0_s \equiv n+1$.

\item \textbf{Step t (t $\geq$ 1) with Sequential Cut-offs:} If each institution $s$ has filled its capacity or has announced the lowest cut-off possible ($c_s=1$), the algorithm stops and returns $\psi[\gamma]^{t-1}$. That is, tentative assignments become permanent. 

A \textbf{single} institution $s$ with $c^{t-1}_{s}>1$ that has not filled its capacity, announces the lowest possible cut-off $c^t_s \in \mathbb{N}$ such that  $|\{i\in I: c^t_s \leq m_s(i) < c^{t-1}_s\}| \leq q_s - | \psi[\gamma]^{t-1}(s)| $. Other institutions announce $c^{t}_{s'} = c^{t-1}_{s'}$.

Let $B^t_i \equiv \{s\in S: c^t_s \leq m_s(i) < c^{t-1}_s \}\cup \psi[\gamma]^{t-1}(i) $ denote the set of institutions that individual $i$ can choose from. Individual $i$'s tentative assignment is updated as follows:

\begin{equation*}
    \psi[\gamma]^{t}(i)= 
\begin{cases}
    C_i(B^t_i) ,& \text{if } C_i(B^t_i )\neq \emptyset\\
    \psi[\gamma]^{t-1}(i),           & \text{otherwise}
\end{cases}
\end{equation*}

Institution $s$'s assignment is $\psi[\gamma]^{t}(s)=\{i \in I: \psi[\gamma]^t(i)=\{s\}\}$. 

\end{itemize}

The algorithm terminates in a finite number of steps. Let $\psi^{seq}$ be the mechanism based on the sequential cut-offs algorithm. The mechanism is not equipped to deal with indecisive choice behavior. In addition to acceptable-consistency, this mechanism requires acyclic choice functions. Formally, a choice function $C_i$ is \textbf{acyclic (over acceptable institutions)} if for all sequences of acceptable institutions $s^1, s^2, \dots , s^t\in S$,
\[C_i(\{s^1,s^2\}) = \{s^1\}, \dots, C_i(\{s^{t-1},s^{t}\}) = \{s^{t-1}\}  \text{ implies } C_i(\{s^{1},s^{t}\}) = \{s^{1}\}.\]

Acyclicity is a stronger requirement than weak acyclicity. Acyclicity rules out weak and strict cycles in binary menus, while weak acyclicity rules out only strict cycles in binary menus. Nonetheless, acyclic and acceptable-consistent choice functions are more general than path independent ones.  Specifically, the proof of \Cref{PlottImpliesWAAC} shows that path independent choice functions are acyclic and acceptable-consistent, meanwhile \Cref{Example: AttractionEffect} shows that acyclic and acceptable-consistent choice functions may not be path independent.
Our final result shows that the sequential cut-offs mechanism requires much less choice sophistication than its simultaneous counterpart.

\begin{prop}\label{Application_Sequential}
The sequential cut-offs mechanism $\psi^{seq}$ leads to a stable outcome $\psi^{seq}[\gamma]$ for every admissions problem  $ \gamma \in \Gamma$ if and only if the choice functions are acyclic and acceptable-consistent.
\end{prop}

\section{Discussion}
\label{section:discussion}
In this section, we discuss several questions that came up in multiple discussions, acknowledging their potential reoccurrence in the mind of a curious reader.

\vspace{-0.3cm}
\paragraph{Single-valued versus multi-valued choice models.}
We model choice as single-valued as opposed to multi-valued.
Multi-valued choice models do not allow for empty choices, and an observation like $C_i(\{s^1,s^2 ,s^3\})=\{s^1,s^2\}$  results from indecisiveness or indifference between the two alternatives (\cite{eliaz2006indifference}). The individual deems the two alternatives equally choosable, selecting $s^1$ sometimes and $s^2$ the other times.
Single-valued choice models are more suited for modeling matching markets, like school choice, university admission, daycare assignment, etc., as participants' choices are observed once, not multiple times. For example, it is unclear how information from a multi-valued choice model should be used when running a dynamic deferred acceptance style mechanism that requires individuals to make a single selection during any step of the procedure.

\vspace{-0.3cm}
\paragraph{Interpreting choices.}  
We interpret (1) choices from singleton menus as revealing whether a given institution is acceptable or unacceptable, and (2) choices from binary or larger menus as revealing preferred institutions, indecisiveness, and indifference. Specifically, $C_i(\{s\})=\{s\}$ means that $s$ is revealed to be preferred to the $\emptyset$ ($s$ is acceptable), and $C_i(\{s\})=\emptyset$ that $\emptyset$ is preferred to $s$ ($s$ is unacceptable). On the other hand,  $C_i(\{s,s'\})=\{s\}$ reveals that $s'$ is preferred over $s$.
Finally, indecisiveness or indifference between $s$ and $s'$ is observed as  $C_i(\{s,s'\})=\emptyset$. For any larger menu, $S'$ with $|S'|>2$, $C_i(S')=\emptyset$ reveals indecisiveness or indifference, but not between which institutions the individual cannot decide.

\vspace{-0.3cm}
\paragraph{Can empty choice be formulated as an institution?}

One of the properties that characterize the standard model of choice taught in graduate schools is that choice from finite menus is non-empty (\cite{kreps2013microeconomic}).
However, in the non-standard choice paradigm, empty choices cannot be ignored, be it due to indecisiveness, indifference,  or choice avoidance (\cite{clark1995indecisive}, \cite{gerasimou2018indecisiveness}). In such context, formulating empty choice as an institution and restricting choice to be non-empty is akin to forcing an individual to choose ``\textit{between inherently incomparable alternatives}'' that may generally lead more inconsistent choices than non-forced choices (\cite{luce1989games}).

\vspace{-0.3cm}
\paragraph{Preference reporting language.}

Our setup raises a question about preference reporting language. 
Analogous to the problem of reporting combinatorial preferences (\cite{milgrom2011critical}, \cite{budish2021can}), eliciting sufficient choice information required for running a direct mechanism is infeasible in practice. 
Dynamic mechanisms offer a simple alternative to direct mechanisms by reducing the need to gather choice information to a minimum (\Cref{section: Application}). By making choices directly, rather than pre-reporting choice functions to a central authority, one can avoid the complexities of defining a preference-reporting language that accurately captures individual choice behavior. This approach comes at the cost of being unable to account for indecisiveness and indifference. This seems to be a promising direction that warrants further investigation.

\vspace{-0.3cm}
\paragraph{Appropriateness of concepts.}
A common motivation for pairwise stable outcomes also applies in our setting. After an initial matching is made, individuals can explore alternative options by searching around and approaching other institutions one by one. In such a scenario, pairwise stability ensures that no pair of individuals and institutions would prefer to leave their current assignment and form a new one. 
The primary motivation behind our solution concepts is their emphasis on the role of binary comparisons, implying that individuals' choices over pairs of options are ``more meaningful" than choices from larger menus, which has an intuitive appeal. 
This is, for example, illustrated by the attraction effect, where an individual's choice is distorted by an irrelevant alternative \citep{huber1982adding}. Here, our pairwise stability notion assigns the best alternative to the individual, absent any distorting irrelevant alternative. We interpret this as guiding individuals exhibiting non-standard choice, to avoid choice mistakes.

While misreporting, individuals focus only on two potential assignments: The first is achieved by truthfully reporting their choices, and the second is based on what can be achieved by misreporting choices. This has some interesting implications. For instance, our notion of incentive compatibility suggests that serial dictatorships may not be incentive compatible due to choice mistakes.\footnote{This is clear from \Cref{Example: AttractionEffect} where the individual would choose $s^2$ out of $\{s^1,s^2,s^3\}$ as the first dictator under serial dictatorship. Meanwhile, misreporting $s^1$ as her choice out of $\{s^1,s^2,s^3\}$ would lead to a pairwise preferred outcome as $s^1$ is her choice out of $\{s^1,s^2\}$.} Instead, if a designer wants to implement a serial dictatorship mechanism, individuals must be guided pairwise through the available options to achieve incentive compatibility. 
Finally, our efficiency concept aligns with the typical notion of Pareto improvement. It focuses on individuals who clearly decide between two options, avoiding empty choices when comparing various assignments. In scenarios characterized by excessive indecisiveness, an efficient outcome might not be attainable because individuals are themselves unsure of their preferred options, as illustrated in \Cref{Example:ConstrainedCycle}.

\vspace{-0.3cm}
\paragraph{Alternative Concepts.}

The richness introduced by moving from binary (preference) relations to choice functions remains largely unexplored due to our solution concepts emphasizing binary comparisons. 
A designer who wants to give more weight to individuals' choices from larger menus, might therefore be interested in solution concepts that are defined on the basis of individuals' choices from some particular reference menu. 
For instance, one could define stability based on the menu of \textit{attainable assignments}, \textit{acceptable assignments}, or \textit{all assignments}.
If we are to broaden the reach of institution design to settings in which agents have unorthodox but realistic patterns of preferences, such approaches to adapting and redefining key concepts are worth scrutinizing further.

\section{Conclusion}

We extend matching theory to problems where individuals may exhibit a plethora of non-standard choice behaviors. We show that weak acyclic and acceptable-consistent choice functions are both necessary and sufficient for the existence of stable matchings and a large class of incentive compatible mechanisms. Compared to the standard choice behavior, characterized by path independent choice functions, our identified conditions allow for more general choice behavior of individuals. Doing so not only affects how one finds stable matchings but also their underlying structure. Finally, we show that a commonly used dynamic mechanism can be tailored to accommodate non-standard choice behavior. 

\pagebreak

\include{mybib.bib}
\bibliographystyle{aer}
\bibliography{mybib}

\newpage

\begin{center} \Large

\end{center}

\appendix
\section{Mathematical Appendix and Proofs} \label{section: AppendixProofs}

\subsection*{\Cref{Stability_Lemma}}

\begin{proof}
Consider a subset $S'\subseteq S$ and individual  $i\in I$ with at least one acceptable institution, that is,  $s\in S'$ such that $C_i(\{s\})=\{s\}$. Moreover, suppose that $U_i(S')=\emptyset$. Define the set of acceptable institutions as  $\overline{S'}=\{s \in S': C_i(\{s\})=\{s\}\} \neq \emptyset$ and the set of unacceptable institutions as $\underline{S'}=\{s \in S': C_i(\{s\})=\emptyset\}$. 

  First, suppose there exists an unacceptable institution  $s'\in \underline{S'}$ that is chosen over an acceptable institution  $s\in \overline{S'}$, i.e., $C_i(\{s,s'\})=\{s'\}$.
Since choice functions are acceptable-consistent we have that $C_i(\{s\})=\{s\}$ and $C_i(\{s'\})=\emptyset$ implies $C_i(\{s,s'\})\neq\{s'\}$ --- a contradiction. Therefore, only an acceptable institution can be chosen over another acceptable institution.

Second, suppose that $|\overline{S'}|\leq 2$. Then there trivially exists a C-maximal institution, as only an acceptable institution can be chosen over another acceptable institution.

Third, suppose that for every acceptable institution $s\in \overline{S'}$, there is another institution $s'\in S'$ that is chosen in pairwise comparison.
By the first part, only an acceptable institution can be chosen over another acceptable institution. That is, for all  $s\in \overline{S'}$ there exists $s'\in \overline{S'}\setminus \{s\}$ such that $C_i(\{s,s'\})=\{s'\}$.
By the second part, $|\overline{S'}|\geq 3$. But this implies that there exists a positive integer $t \geq 3$ and $t$ distinct and acceptable alternatives $s^1, s^2, \dots , s^t$, such that $C_i(\{s^1,s^2\}) = \{s^2\}, \dots, C_i(\{s^{t-1},s^{t}\}) = \{s^{t}\}, \text{ and } C_i(\{s^{t},s^{1}\}) = \{s^{1}\}$ ---  a contradiction to weak acyclicity. 
\end{proof}

\subsection*{\Cref{Stability_MainThm}}

We construct an outcome for every admission problem $\gamma \in \Gamma$ using the algorithm described next. 

\bigskip

\noindent \textbf{Non-Block Algorithm}

\noindent \textbf{Step 0:} For each $i\in I$ consider the set of  mutually acceptable institutions $S^0_i\equiv \{s\in S: C_i(\{s\})=\{s\} \text{ and } s \pi_s \emptyset \}$.
Let the set of individuals proposing to institution $s$ be denoted by $I^0_s\equiv \emptyset$.
Finally, let $A_s(I')\equiv \{i\in I': |\{i'\in I': i' \pi_s i\}|< q_s\}$ denote the set of individuals in $I'\subseteq I$ tentatively assigned to institution $s$. 

\medskip
\noindent \textbf{Step $\pmb{k\geq 1}$:} 
If there is at least one individual currently not tentatively admitted, i.e., $i \not \in \bigcup_{s\in S} A_s(I^{k-1}_s)$, and that still has a mutually acceptable institution left to propose, i.e., $S^{k-1}_i\neq \emptyset$, then let each such individual $i$ propose to an C-maximal institution $s^k_i\in S^{k-1}_i$ (if there are multiple, take the lowest-subscript institution) --- by \Cref{Stability_Lemma} such an institution exists since $S^{k-1}_i\neq \emptyset$ and contains only acceptable institutions. Let $P^k_s$ denote the set of individuals proposing to institution $s$ at step $k$.
Set $S^k_i=S^{k-1}_i\setminus \{s^k_i\}$ and $I^k_{s}=I^{k-1}_s \cup P^k_s$. Go to step $k+1$.

\medskip
 \noindent Otherwise, if no such individual exists, the algorithm stops. Then each institution is assigned $A_s(I^{k-1}_s)$ while each individual $i$ is assigned $s\in S$ such that $i\in A_s(I^{k-1}_s)$ and $\emptyset$ otherwise.

\bigskip

 Let the \textbf{non-block mechanism $\psi^{nb}$} be described by the function that associates the outcomes of the non-block algorithm to any admission problem $\gamma \in \Gamma$. We now prove \Cref{Stability_MainThm}.

\begin{proof}

\textbf{The ``if" part:} If the choice functions $(C_i)_{i\in I}$ are weakly acyclic and acceptable-consistent, then the non-block mechanism $\psi^{nb}$ yields a stable matching $\psi^{nb}[\gamma]$ for every admission problem $\gamma \in \Gamma$.

For any admission problem $\pi \in \Pi$, consider the outcome of the non-block mechanism $\psi^{nb}[\pi]$. Since individuals propose to mutually acceptable institutions only, the outcome $\psi^{nb}[\pi]$ is trivially individually rational for both institutions and individuals.

Next, suppose that there exists $s\in S\setminus \psi^{nb}[\pi](i)$ with $C_i(\{s\}\cup \psi[\pi](i))=\{s\}$ and $s\pi_s \emptyset$.
By acceptable-consistency, $s$ must be acceptable. Together with \Cref{Stability_Lemma}, this implies that $i$ must have proposed to $s$ at some step $k$ and subsequently been rejected. Therefore, for institution $s$ it must be the case that $|\psi^{nb}[\pi](s)|=q_s$ and $i'\pi_s i$ for all  $i'\in \mu[\pi](s)$.

\bigskip

\noindent \textbf{The ``only if" part:} A stable matching $\mu$ exists for every admission problems $\gamma \in \Gamma$ only if the choice functions $(C_i)_{i\in I}$ are weakly acyclic and acceptable-consistent. The contrapositive statement is, if the choice functions $(C_i)_{i\in I}$ are not weakly acyclic or acceptable-consistent, then for some admissions problem $\gamma\in \Gamma$ a stable matching $\mu$ does not exist.

\bigskip

\noindent \textbf{Part 1. Weak Acyclicity is necessary.}\\ Suppose that for some $C_i$ we have a cycle for $t\geq 3$ and distinct acceptable alternatives $s^1, s^2, \dots , s^t$ in $S$ such that  $C_i(\{s^1,s^2\}) = \{s^2\}, \dots C_i(\{s^{t-1},s^{t}\}) = \{s^{t}\}, \text{ and } C_i(\{s^{t},s^{1}\}) = \{s^{1}\}$.
 
Consider an admissions problem $\tilde{\gamma} =\langle I, S, q, C, \tilde{\pi} \rangle$ such that $i \mathrel{\tilde{\pi}}_s i'$ for all $i' \in I\setminus \{i\}$ and $s \in \{ s^1, s^2, \dots , s^t\}$ and $\emptyset \mathrel{\tilde{\pi}}_s i$ for all $s \in S \setminus \{ s^1, s^2, \dots , s^t\}$.  Consider a matching $\mu$ for problem $\tilde{\gamma}$. If $\mu(i)=\{s\}$ for some $s\in \{s^1, s^2, \dots , s^t\}$ or $\mu(i)=\emptyset$, $i$ forms a blocking pair with some $s' \in \{s^1, s^2, \dots , s^t\}\setminus \mu(i) $. While if  $\mu(i)=\{s\}$ for $s\in S\setminus \{s^1, s^2, \dots , s^t\}$ we have a violation of individual rationality for the appropriate institution. Therefore, there is no stable matching for problem $\tilde{\gamma}$.

\bigskip

\noindent \textbf{Part 2. Acceptable-consistency is necessary.}\\ Suppose that for some $C_i$ we have $C_i(\{s^1\})=\{s^1\}$  and $C_i(\{s^2\})=\emptyset$ but $C_i(\{s^1,s^2\})= \{s^2\}$.

Consider an admission problem $\tilde{\gamma} \in \Gamma$ such that $i \mathrel{\tilde{\pi}}_s i'$ for all $i' \in I\setminus \{i\}$ and $s \in \{ s^1,s^2\}$ and $\emptyset \mathrel{\tilde{\pi}}_s i$ for all $s \in S \setminus  \{ s^1,s^2\}$. Consider a matching $\mu$ for problem $\tilde{\gamma}$. There are four possibilities. (i) If $\mu(i)=\{s^1\}$, $i$ forms a blocking pair with $s^2$. (ii) If $\mu(i)=\{s^2\}$ we have a violation of individual rationality for individual $i$. (iii) If $\mu(i)=\emptyset$,  $i$ forms a blocking pair with $s^1$. (iv) While if  $\mu(i)=\{s\}$ for $s\in S\setminus \{s^1,s^2\}$ we have a violation of individual rationality for the appropriate institution. Therefore, there is no stable matching for problem $\tilde{\gamma}$.
\end{proof}

\paragraph{Detour: Why do we not use the Szpilrajn extension theorem to construct a preference from the binary choice of a weakly acyclic and acceptable-consistent choice function $C_i$?} 

Consider the problem of constructing  a binary relation $R_i$ for a weakly acyclic and acceptable-consistent choice function $C_i$ such that  $C_i(\{s^1,s^2\}) =\{ s^1\}$, $C_i(\{s^2,s^3\}) = \{s^2\}$, and $C_i(\{s^1,s^3\}) = \emptyset$. To handle the case where choice out of a binary menu is empty, there are only two possible approaches: either $C_i(\{s^1,s^3\})=\emptyset$ implies both  $s^1 R_i s^3$ and $s^3 R_i s^1$, or neither. The first approach is generally interpreted as an individual being indifferent between $s^1$ and $s^3$. In contrast, the second approach leaves more room for non-standard interpretation, like indecisiveness and choice avoidance. 

Suppose we take the second approach, while the resulting binary relation might look like a partial order (asymmetric and transitive) at first glance, it fails transitivity for the described choice function $C_i$ --- as $s^1 R_i s^2$,  $s^2 R_i s^3$, but not $s^1 R_i s^3$.  Immediately, we cannot apply the Szpilrajn extension theorem, which shows that every partial order is contained in a simple order  (reflexive, transitive, anti-symmetric, and complete). 

If we instead take the first approach,  while the resulting binary relation might look like a simple order at first glance, it is, e.g., not anti-symmetric --- as $s^1 R_i s^3$,  $s^3 R_i s^1$ but $s^1\neq s^3$. 
Therefore, in \Cref{ProxyMarket_Existence} we simply take  the brute force approach of either choosing $P_i$ such that  $s P_i s$ or $s' P_i s$ --- neither one directly reflecting that $C_i(\{s^1,s^3\}) = \emptyset$.
This observation already hints at \Cref{Richness_Stability}; that is, while we can find stable matchings through constructing an associated proxy admission problem, not every stable matching in the original problem might be found in such a way. 

\subsection*{\Cref{PlottAdjusted}}

We start by proving the following lemma.

\begin{lem}
\label{pathindependence}
Consider a path independent choice function $C_i$, then 
\begin{enumerate}[(i)]
\item an unacceptable alternative cannot be chosen, that is,  $$ C_i(\{s\})=\emptyset \implies C_i(\{s\} \cup S')\neq \{s\} \quad \text{for any} \quad  S'\subseteq S\setminus \{s\}.$$
\item choice from a set containing at least one acceptable alternative is non-empty, that is, 
$$C_i(\{s\})=\{s\} \implies C_i(\{s\} \cup S')\neq \emptyset \quad \text{for any} \quad  S'\subseteq S\setminus \{s\}.$$
\item choice from a set containing only unacceptable alternatives is empty, that is, 
$$C_i(\{s\})=\emptyset \quad \text{for all} \quad s \in S' \subseteq S   \implies C_i(S')=\emptyset.$$
\end{enumerate}
\end{lem}

\begin{proof}
(i) Consider $s\in S$ such that $C_i(\{s\})=\emptyset$ and any $S'\subseteq S\setminus \{s\}$. By path independence we have
\begin{align*}
 C_i(\{s\} \cup S') &= C_i(C_i(\{s\})\cup S')\\
                    &=C_i(S')\subseteq S'\\
                    &\neq \{s\}.
\end{align*}

\noindent (ii) Consider $s\in S$ such that $C_i(\{s\})=\{s\}$ and any $S'\subseteq S\setminus \{s\}$. Assume towards a contradiction that $C_{i}\left(S^{\prime} \cup\{s\}\right)=\emptyset$. By path independence  we have
\begin{align*}
    C_i(\{s\} \cup S')&= C_i(\{s\}\cup (\{s\} \cup S'))\\
    &=C_i(\{s\}\cup C_i( S'\cup \{s\}))\\
    &=C_i(\{s\})\\
    &=\{s\}\\
    &\neq \emptyset.
\end{align*}

\noindent (iii) Consider any $S'\subseteq S$ with  $C_i(\{s\})=\emptyset$ for all $s \in S'$. 
Let $S'=\{s_1,\dotso ,s_k\}$. By path independence we can remove alternatives one by one to reach the desired conclusion, that is,
\begin{align*}
    C_i(S') &= C_i(C_i(\{s_1\})\cup (S'\setminus \{s_1\}))\\
    &=C_i(S'\setminus \{s_1\})\\
    &=C_i(C_i(\{s_2\})\cup (S'\setminus \{s_1,s_2\}))\\
    &=\dots\\
    &=C_i(\{s_k\})\\
    &=\emptyset.
\end{align*}

\end{proof}

\noindent We now prove \Cref{PlottAdjusted}.

\begin{proof}
\textbf{The ``if" part:} If a unit demand choice function $C_i$ is path independent then it is rationalizable by a simple order $R_i$ over acceptable choices.

Consider $C_i$ such that $C_i(\{s,s'\})=\{s\}$ and $C_i(\{s',s''\})=\{s'\}$. Note that path independence implies that
\begin{align*}									
\{s\} 		&=C_i(\{s\} \cup \{s'\}) &   \\
		&=C_i(\{s\} \cup C_i(\{s',s''\})) & \\
		&=C_i(\{s\} \cup \{s',s''\}) & 	 \text{ by path independence} \\
	&=C_i(C_i(\{s, s'\}) \cup \{s''\}) & \text{ by path independence} \\
	&=C_i(\{s,s''\}).
\end{align*} 

 Next, define binary relation $P_i$ such that $s \mathrel{P_i} s'$ for $s,s'\in S$ with $s\neq s'$ if $C_i(\{s,s'\})=\{s\}$, $C_i(\{s\})=\{s\}$, and $C_i(\{s'\})=\{s'\}$.
Moreover, $s \mathrel{P_i} \emptyset$ if $C_i(\{s\})=\{s\}$, and $\emptyset \mathrel{P_i} s$ if $C_i(\{s\})=\emptyset$.

Similarly, define $R_i$ such that $s R_i s'$ if and only if $[s \mathrel{P_i} s' \text{ or } s=s']$, $s R_i \emptyset$ if and only if $s \mathrel{P_i} \emptyset$, and $\emptyset R_i s$ if and only if $\emptyset \mathrel{P_i} s$. Note that,  $R_i$ is (strongly) complete over acceptable choices. This is because for two distinct acceptable $s, s'$ we always have $s \mathrel{P_i} s'$ or $s' \mathrel{P_i} s$ as $C_i(\{s,s'\})$ is non-empty by \Cref{pathindependence} part (ii). $R_i$  is anti-symmetric by construction. Moreover,  $R_i$  is transitive over acceptable choices as for path independent choice functions we have established above that $C_i(\{s,s'\})=\{s\}$ and $C_i(\{s',s''\})=\{s'\}$ imply $C_i(\{s,s''\})=\{s\}$. For the constructed order $R_i$ we next show that 
\begin{enumerate}[(i)]
    \item $C_i(S')=\emptyset$ if $\emptyset R_i s$ for all $s\in S'$, and
    \item $C_i(S')=\{s\in S': s R_i s' \text{ for all } s'\in S' \}$  otherwise.
\end{enumerate}

(i) Consider $S'\subseteq S$ such that $\emptyset R_i s$ for all $s\in S'$. For any $s\in S'$, by construction $\emptyset R_i s$ implies $\emptyset \mathrel{P_i} s$ which implies $C_i(\{s\})=\emptyset$.
By \Cref{pathindependence}, if $C_i(\{s\})=\emptyset$ for all $s\in S'$ then $C_i(S')=\emptyset$.

(ii) Consider $S'\subseteq S$ with at least one $s\in S'$ such that $sR_i \emptyset$. By construction $s R_i \emptyset$ implies $s\mathrel{P_i} \emptyset$ which implies $C_i(\{s\})=\{s\}$.
That is, $S'$ contains  at least one acceptable alternative. Suppose by contradiction that $C_i(S')\neq \{s\in S': sR_i s' \text{ for all } s'\in S'\}$. By \Cref{pathindependence}, $C_i(S')\neq \emptyset$.
Hence, let $C_i(S')=\{s'\}$ and  $\{s\in S': sR_i s' \text{ for all } s'\in S'\}=\{s\}$.
Since $s\neq s'$ and $sR_i s'$ we have $s\mathrel{P_i} s'$ respectively  $\{s\}=C_i(\{s,s'\})$.
From this observation we reach a contradiction since
\begin{align*}									
\{s'\} 		&=C_i(\{s,s'\} \cup (S' \setminus \{s,s'\})) &   \\
&=C_i(\{s\} \cup (S' \setminus \{s,s'\}) &  \text{ by path independence} \\
&= C_i(S' \setminus \{s'\})  &\\
&\neq \{s'\}.
\end{align*}

\bigskip

\noindent \textbf{The ``only if" part:} If a unit demand choice function $C_i$  is rationalizable by a simple order $R_i$ then it is path independent. 

\noindent There are three cases to consider.

\begin{enumerate}[(i)]
    \item Consider $S', S'' \subseteq S$  both containing at least one acceptable alternative, that is, $s$ such that $s R_i \emptyset$. Let $\{s^*\}=\{s\in S'\cup S'': sR_i s' \text{ for all } s'\in S'\cup S''\}$, and $\{s^{*\prime}\}=\{s\in S': sR_i s' \text{ for all } s'\in S'\}$.
Since $s^* R_i s' \text{ for all } s'\in S'\cup S''$ and $s^{*\prime}\in S'$ we have $\{s^*\}=\{s\in \{s^{*\prime}\}\cup S'': sR_i s' \text{ for all } s'\in \{s^{*\prime}\}\cup S''\}$. Rewriting this in terms of choice functions leads 
\begin{align*}									
C_i(S'\cup S'')&=C_i(\{s^{*\prime}\}\cup S'') &   \\
&=C_i(C_i(S')\cup S'') & 
\end{align*}

 \item  Consider $S', S'' \subseteq S$  both containing no acceptable alternatives. We have $C(S')=\emptyset$, $C(S'')=\emptyset$ and $C(S'\cup S'')=\emptyset$. It directly follows that $C_i(S'\cup S'')=C_i(S'') =C_i(C_i(S')\cup S'')$.

 \item  Consider $S', S'' \subseteq S$ where only $S'$ contains at least one acceptable alternative. 
Let $\{s^*\}=\{s\in S'\cup S'': sR_i s' \text{ for all } s'\in S'\cup S''\}$. It follows that $\{s^*\}=\{s\in S': sR_i s' \text{ for all } s'\in S'\}$ as well as $\{s^*\}=\{s\in S''\cup\{s^*\}: sR_i s' \text{ for all } s'\in S''\cup\{s^*\}\}$. Again, we get that $C_i(S'\cup S'')=C_i(C_i(S')\cup S'')$.
\end{enumerate}
\end{proof}

\subsection*{\Cref{PlottImpliesWAAC}}

\begin{proof}
\textbf{Part 1. Path independent choice functions are weakly acyclic.}\\

Consider a path independent choice function $C_i$, an integer $t \geq 3$ and $t$ distinct and acceptable institutions $s^1, s^2, \dots , s^t\in S$,
$C_i(\{s^1,s^2\}) = \{s^1\}, \dots, C_i(\{s^{t-1},s^{t}\}) = \{s^{t-1}\}$.

First, notice that path independence implies that
\begin{align*}									
C_i(\{s^1, s^2, \dots , s^t\}) 		&=C_i(\{s^1, s^2, \dots , s^{t-2}\} \cup C_i(\{s^{t-1}, s^{t}\})) &   \\
&=C_i(\{s^1, s^2, \dots , s^{t-3}\} \cup C_i(\{s^{t-2}, s^{t-1}\})) &   \\
		&=\dots & \\
		&= \{s^1\}\\
  &\neq \{s^t\}.
\end{align*}

Second, path independence implies that
\begin{align*}									
C_i(\{s^1, s^2, \dots , s^t\}) 		&=C_i(\{s^1, s^2, \dots , s^{t-3}\} \cup C_i(\{s^{t-2}, s^{t-1}\} )\cup \{s^t\}) &   \\
		&=C_i(\{s^1, s^2, \dots , s^{t-4}\} \cup C_i(\{s^{t-3}, s^{t-2}\}) \cup \{s^t\} )& \\
		&= \dots \\ 
		&= C_i(\{s^1, s^t\}).
\end{align*}
 Therefore, $C_i(\{s^1, s^t\}) \neq \{s^t\}$, that is, $C_i$ is weakly acyclic. 

\noindent Remark: As $C_i(\{s^1, s^t\}) = \{s^1\}$, we have also shown that, path independent $C_i$ is acyclic (as defined in \Cref{section: Application}). 

\bigskip

\noindent \textbf{Part 2. Path independent choice functions are acceptable-consistent.}\\
Consider a path independent choice function $C_i$ and distinct institutions $s,s'\in S$ such that 
$C_i(\{s\})=\{s\}$  and $C_i(\{s'\})=\emptyset$. Path independence implies that 
\begin{align*}									
C_i(\{s,s'\}) 		&=C_i(\{s\} \cup C_i\{s'\} ) &   \\
		&=C_i(\{s\}) &\\
		& \neq \{s'\}.
\end{align*}
Therefore, $C_i$ is acceptable-consistent.

\bigskip

\noindent \textbf{Part 3. Weakly acyclic and acceptable-consistent choice functions may not be path independent.}

See \Cref{Example: AttractionEffect}.

\end{proof}

\subsection*{\Cref{ProxyMarket_Existence}}

\begin{proof}
Consider the following construction for each individual $i\in I$:

\noindent \textbf{Step 0:}
First, consider the unacceptable assignments, that is, $\{s\in S: C_i(\{s\})= \emptyset\}$ and order them as $\emptyset P_i s$. Second, consider the acceptable assignments $S^0=S\setminus \{s\in S: C_i(\{s\})= \emptyset\}$. If $S^0= \emptyset$ the construction stops. Otherwise consider the set of C-maximal institutions  $U(S^0)=\{s\in S^0: \{C_i(\{s,s'\})=\{s'\} \text{ for some } s'\in S\setminus \{s\} \}=\emptyset \}$ --- which is non-empty by \Cref{Stability_Lemma}. Let
\begin{enumerate}[(i)]
    \item $sP_is'$ for all $s\in U(S^0)$ and $s' \in S^0\setminus U(S^0)$,

   \item order elements of $U(S^0)$ such that
     \begin{itemize}
         \item for all $s,s' \in U(S^0)$ with $s\neq s'$ either $s P_i s'$ or  $s' P_i s$, 
         \item for all $s,s' \in U(S^0)$ with  $s P_i s'$ implies $\neg (s' P_i s)$, and
         \item for all $s,s',s'' \in U(S^0)$ with $sP_i s'$ and $s' P_i s''$ implies $s P_i s''$. 
     \end{itemize}
\end{enumerate}

\medskip

\noindent \textbf{Step $\pmb{k\geq 1}$:} 
Consider $S^k=S^{k-1}\setminus U(S^{k-1})$. If $S^k= \emptyset$ the construction stops. Otherwise consider the set of C-maximal institutions $U(S^k)$ ---- which is non-empty by \Cref{Stability_Lemma}. Let
\begin{enumerate}[(i)]
    \item $sP_is'$ for all $s\in U(S^k)$ and $s' \in S^{k}\setminus U(S^k)$,
   \item order elements of $U(S^k)$ such that
     \begin{itemize}
         \item for all $s,s' \in U(S^k)$ with $s\neq s'$ either $s P_i s'$ or  $s' P_i s$, 
         \item for all $s,s' \in U(S^k)$ with  $s P_i s'$ implies $\neg (s' P_i s)$, and
         \item for all $s,s',s'' \in U(S^k)$ with $sP_i s'$ and $s' P_i s''$ implies $s P_i s''$. 
     \end{itemize}
\end{enumerate}

\bigskip

\noindent This construction yields a strict simple order over acceptable choices satisfying the conditions that (1) if $C_i(\{s,s'\})=\{s\}$ then $s \mathrel{P_i} s'$; (2) if $C_i(\{s\})=\{s\}$ then $s \mathrel{P_i} \emptyset$; and, (3) if $C_i(\{s\})=\emptyset$ then $\emptyset \mathrel{P_i} s$. 
Constructing $P_i$ for all $i\in I$ we end up with $P=(P_i)_{i\in I}$ giving us the proxy admission problem $\gamma_P = \langle I, S, q, P, \pi\rangle$. Thus, the set of proxy admission problems $\Gamma_{\gamma}$ is non-empty. 

\end{proof}

\subsection*{\Cref{Richness_Stability}}
\begin{proof}
Fix an admissions problem $\Gamma$ with choice functions that are weakly acyclic and acceptable-consistent.

\bigskip

\noindent \textbf{The ``if" part:}  If  a  matching  is  stable  for  some  associated  proxy  admissions  problem  then  it  is  also stable for the admissions problem.

Consider a one-to-one mapping $f: \Gamma \mapsto \Gamma_\mathcal{P}$ such that $f(\gamma) \in \Gamma_{\gamma}$ --- which is non-empty by \Cref{ProxyMarket_Existence}. Let $\hat{\psi}^s$ denote a stable mechanism for proxy admissions problems. Now consider the mechanism $\psi^{s}$ for admissions problems such that 
$\psi^{s}[\gamma]\equiv \hat{\psi}^{s}[f(\gamma)]$ for all $\gamma \in \Gamma$. 

\begin{enumerate}[(i)]
\item  Suppose $\psi^{s}[\gamma]$ is not individually rational for some $\gamma \in \Gamma$, then there exists $i\in I$ with $\psi^{s}[\gamma](i) \neq \emptyset$ such that  $C_i(\psi^{s}[\gamma](i))=\emptyset$. This implies that in the associated proxy admissions problem $\emptyset \mathrel{P}_i \hat{\psi}^{s}[f(\gamma)](i)$. Thus contradicting that $\hat{\psi^{s}}$ is stable for proxy admissions problems.

\item  Suppose there exist $\gamma \in \Gamma$ such that   $\psi^{s}[\gamma]$ has a blocking pair $(i,s)$, i.e., $C_i(\psi^{s}[\gamma](i) \cup {s})={s}$ and $i\pi_s i'$ for some $i'\in \psi^{s}[\gamma](s)$, or $|\psi^{s}[\gamma](s)|<q_s$ with $i\pi_s \emptyset$. Note that, $C_i(\psi^{s}[\gamma](i) \cup {s})={s}$ implies that in the associated proxy admissions problem  $s \mathrel{P_i} \hat{\psi}^{s}[f(\gamma)]$. Thus contradicting that $\hat{\psi}^{s}$ is stable for proxy admissions problems.
\end{enumerate}

\bigskip

\noindent \textbf{The converse may not hold:} A stable matching for the admissions problem may not be stable  for any  associated  proxy  admissions  problem in $\Gamma_\gamma$.

See \Cref{Richness_Example1}.
\end{proof}

\subsection*{\Cref{Richness_Stability2}}
\begin{proof}
Fix an admissions problem $\Gamma$ with choice functions that are path independent.

\bigskip

\noindent \textbf{The ``if" part:}  If  a  matching  is  stable  for  some  associated  proxy  admissions  problem  then  it  is  also stable for the admissions problem.

Since path independent choice functions are weakly acyclic and acceptable-consistent (\Cref{PlottImpliesWAAC}), this part is a corollary to \Cref{Richness_Stability}.

\bigskip

\noindent \textbf{The ``only if" part:} A stable matching for the admissions problem is also stable  for some  associated  proxy  admissions  problem in $\Gamma_\gamma$.

Notice that with path independent choice functions there is a unique proxy admissions problem for every admissions problem with the preferences constructed in the same way as in \Cref{PlottAdjusted}. 

Consider a one-to-one mapping $f: \Gamma \mapsto \Gamma_\mathcal{P}$ such that $f(\gamma) \in \Gamma_{\gamma}$ --- where $P$ the simple order over acceptable choices is constructed in the same way as in \Cref{PlottAdjusted}. Let ${\psi}^s$ denote a stable mechanism for admissions problems. Now consider the mechanism $\hat{\psi}^{s}$ for the proxy admissions problems such that 
$\hat{\psi}^{s}[f(\gamma)] \equiv {\psi}^{s}[\gamma]$ for all $\gamma \in \Gamma$. 

\begin{enumerate}[(i)]
\item  Suppose $\hat{\psi}^{s}[f(\gamma)]$ is not individually rational for some $\gamma \in \Gamma$, then there exists $i\in I$ with $\hat{\psi}^{s}[f(\gamma)](i) \neq \emptyset$ such that $\emptyset \mathrel{P}_i \hat{\psi}^{s}[f(\gamma)](i)$ . This implies that in the associated admissions problem $C_i({\psi}^{s}[\gamma](i))=\emptyset$. Thus contradicting that ${\psi^{s}}$ is stable for proxy admissions problems.

\item  Suppose there exist $\gamma \in \Gamma$ such that   $\hat{\psi}^{s}[f(\gamma)]$ has a blocking pair $(i,s)$, that is, $s \mathrel{P_i} \hat{\psi}^{s}[f(\gamma)]$ and $i\pi_s i'$ for some $i'\in \hat{\psi}^{s}[f(\gamma)](s)$, or $|\hat{\psi}^{s}[f(\gamma)](s)|<q_s$ with $i\pi_s \emptyset$. Note that, $s \mathrel{P_i} \hat{\psi}^{s}[f(\gamma)]$ implies that in the associated admissions problem  $C_i(\psi^{s}[\gamma](i) \cup {s})={s}$ . Thus contradicting that ${\psi}^{s}$ is stable for proxy admissions problems.
\end{enumerate}
\end{proof}

\subsection*{\Cref{Incentives_MainThm} }

We start by defining a related proxy admission problem for any assignment problem, as well as the stability and strategy-proofness in the proxy admission problem. In a second step, we will use the connection between the two problems to prove our result. 

Recall the \textbf{proxy admissions problem} $\gamma_P=\langle I, S, q, P, \pi \rangle$ for admissions problem $\gamma=\langle I, S, q, C, \pi \rangle$, where $P$ is a profile of strict simple orders over acceptable choices satisfying the following conditions for each choice function: 
\begin{enumerate}[(1)]
    \item If $C_i(\{s,s'\})=\{s\}$ then $s \mathrel{P_i} s'$;
    \item If $C_i(\{s\})=\{s\}$ then $s \mathrel{P_i} \emptyset$; and,
    \item If $C_i(\{s\})=\emptyset$ then $\emptyset \mathrel{P_i} s$.
\end{enumerate}
We denote the set of all proxy admission problems by $\Gamma_\mathcal{P}$.

A \textbf{mechanism for proxy admissions problems} is a function $\hat{\psi}: \Gamma_\mathcal{P} \rightarrow \mathcal{M}$ that assigns a matching $\hat{\psi}[\gamma_P] \in \mathcal{M}$ to each proxy admission problem $\gamma_P \in \Gamma_\mathcal{P}$. A mechanism $\hat{\psi}$ is said to be \textbf{incentive compatible (for individuals) for proxy admissions problems} if for any $\gamma_P=\langle I, S, q, P, \pi \rangle$ there does not exist $\gamma_{\hat{P}}=\langle I, S, q, (\hat{P}_i, P_{-i}), \pi \rangle$  such that
\[
\hat{\psi}[\gamma_{\hat{P}}](i) \mathrel{P_i}  \hat{\psi}[\gamma_P](i).\]

A matching $\mu$ is \textbf{stable for proxy admissions problems} $\gamma_P$ if
\begin{enumerate}[(1)]
\item it is \textbf{individually rational}, that is, there is no individual $i$ such that $\emptyset \mathrel{P_i} \mu(i)$ and no institution $s$ such that $\emptyset \mathrel{\pi_s} i$ for some $i\in \mu(s)$, and 
\item there is no \textbf{blocking pair}, that is, there is pair $(i,s)\in I\times S$ such that
 \begin{enumerate}
 \item $s \mathrel{P_i} \mu(i)$, and
 \item 
 	\begin{enumerate}[(i)]
 	\item either  $i\mathrel{\pi_s} i'$ for some $i'\in \mu(s)$, or 
 	\item $|\mu(s)|<q_s$ and $i\mathrel{\pi_s} \emptyset$.
	\end{enumerate}
 \end{enumerate}

\end{enumerate}

A mechanism $\hat{\psi}$ for proxy admissions problems is said to be \textbf{stable for proxy admissions problems} if it assigns a stable matching $\hat{\psi}[\gamma_P]$ to each proxy admission problem $\gamma_P \in \Gamma_\mathcal{P}$. Let $\hat{\psi}^{GS}$ denote the individual-proposing deferred acceptance algorithm defined in \cite{gale1962college}. Recall that this mechanism is both stable and incentive compatible.

\begin{prop}[\cite{gale1962college}]\label{DAstable}
The individual-proposing deferred acceptance mechanism $\hat{\psi}^{GS}$ is stable for proxy admissions problems.
\end{prop}

\begin{prop}[\cite{dubins1981machiavelli}, \cite{Roth1982}]\label{DAstrategyproof}
The individual-proposing deferred acceptance mechanism $\hat{\psi}^{GS}$ is  incentive compatible (for individuals) for proxy admissions problems.
\end{prop}

\noindent We now prove \Cref{Incentives_MainThm}.
\begin{proof} \textbf{Part (a). $1\implies 2$:} Let us show that, if choice functions are weakly acyclic and acceptable-consistent, then we can construct a stable and incentive compatible mechanism. 
Consider a one-to-one mapping $f: \Gamma \mapsto \Gamma_\mathcal{P}$ such that $f(\gamma) \in \Gamma_{\gamma}$ --- which is non-empty by \Cref{ProxyMarket_Existence}. Moreover let $f$ be such that for any $\gamma=\langle I, S, q, C, \pi \rangle$ and $\hat{\gamma}=\langle I, S, q, (\hat{C}_i, C_{-i}), \pi \rangle$ the proxy admissions problem only differ by the preference relation of individual $i$. Such a requirement can be easily accommodated following a construction akin to the one in \Cref{ProxyMarket_Existence}.

Consider the mechanism $\psi^{GS}$ such that 
$\psi^{GS}[\gamma]\equiv \hat{\psi}^{GS}[f(\gamma)]$ for all $\gamma \in \Gamma$. 

\begin{enumerate}[(i)]
\item  Suppose $\psi^{GS}[\gamma]$ is not individually rational for some $\gamma \in \Gamma$, then there exists $i\in I$ with $\psi^{GS}[\gamma](i) \neq \emptyset$ such that  $C_i(\psi^{GS}[\gamma](i))=\emptyset$. This implies that in the associated proxy admissions problem $\emptyset \mathrel{P}_i \hat{\psi}^{GS}[f(\gamma)](i)$. Thus contradicting that $\hat{\psi^{GS}}$ is stable for every proxy admissions problem (\Cref{DAstable}).

\item  Suppose there exist $\gamma \in \Gamma$ such that   $\psi^{GS}[\gamma]$ has a blocking pair $(i,s)$, i.e., $C_i(\psi^{GS}[\gamma](i) \cup {s})={s}$ and $i\pi_s i'$ for some $i'\in \psi^{GS}[\gamma](s)$, or $|\psi^{GS}[\gamma](s)|<q_s$ with $i\pi_s \emptyset$. Note that, $C_i(\psi^{GS}[\gamma](i) \cup {s})={s}$ implies that in the associated proxy admissions problem  $s \mathrel{P_i} \hat{\psi}^{GS}[f(\gamma)]$. Thus contradicting that $\hat{\psi}^{GS}$ is stable for every proxy admissions problem (\Cref{DAstable}).

\item  Suppose $\psi^{GS}$ is not incentive compatible for some $\gamma \in \Gamma$. That is, there exist $\gamma=\langle I, S, q, C, \pi \rangle$ and $\hat{\gamma}=\langle I, S, q, (\hat{C}_i, C_{-i}), \pi \rangle$  such that $
\psi^{GS}[\gamma](i) \neq \psi^{GS}[\hat{\gamma}](i)$ and $ C_i(\psi^{GS}[\hat{\gamma}](i)\cup \psi^{GS}[\gamma](i))= \psi^{GS}[\hat{\gamma}](i)$. 
By construction, in the proxy admissions problem we have that 
$ \hat{\psi}^{GS}[f(\hat{\gamma})](i) \mathrel{P}_i \hat{\psi}^{GS}[f(\gamma)](i)$, thus contradicting that $\hat{\psi}^{GS}$ is incentive compatible (\Cref{DAstrategyproof}).

\end{enumerate}

\bigskip
\noindent \textbf{Part (b). $2\implies 3$:} That is, if there exists a stable and incentive compatible mechanism then there exists a individually rational, weakly non-wasteful, and incentive compatible mechanism.

\noindent This statement trivially holds true as stability implies both individual rationality and weak non-wastefulness.

\bigskip

\noindent \textbf{Part (c). $3\implies 1$:} If there exists an individually rational, weakly non-wasteful, and incentive compatible mechanism then choice functions are weakly acyclic and acceptable-consistent. \\ We prove the contrapositive, i.e., if choice functions are not weakly acyclic or acceptable-consistent, then there exists no individually rational, weakly non-wasteful, and incentive compatible mechanism.

\noindent \textbf{Part 1. Weak acyclicity is necessary.}\\ Suppose that for some $C_i$ we have a cycle for $t\geq 3$ and distinct acceptable alternatives $s^1, s^2, \dots , s^t$ in $S$ such that  $C_i(\{s^1,s^2\}) = \{s^2\}, \dots C_i(\{s^{t-1},s^{t}\}) = \{s^{t}\}, \text{ and } C_i(\{s^{t},s^{1}\}) = \{s^{1}\}$.
 
Consider an admissions problem $\tilde{\gamma} =\langle I, S, q, C, \tilde{\pi} \rangle$ such that $i \mathrel{\tilde{\pi}}_s i'$ for all $i'\in I\setminus \{i\}$ and $s \in \{ s^1, s^2, \dots , s^t\}$ and $\emptyset \mathrel{\tilde{\pi}}_s i$ for all $s \in S \setminus \{ s^1, s^2, \dots , s^t\}$. Moreover, $\tilde{\pi}$ is such that $|\{i'\in I \setminus\{i\}: i' \mathrel{\tilde{\pi}}_s \emptyset \}|< q_s$ for all $s \in \{ s^1, s^2, \dots , s^t\}$ --- which there is at least one such priority order as $q_s\geq 1$. 

If $\psi[\gamma](i)=\{s\}$ for $s\in S\setminus \{s^1, s^2, \dots , s^t\}$ we have a violation of individual rationality for the appropriate institution. Suppose then that the outcome of a mechanism is  $\psi[\gamma](i)=\{s\}$ for some $s\in \{s^1, s^2, \dots , s^t\}$ or $\psi[\gamma](i)=\emptyset$. Consider the following problem $\hat{\gamma}=\langle I, S, q, (\hat{C}_i, C_{-i}), \tilde{\pi} \rangle$  with $\hat{C}_i$ as follows:
\begin{enumerate}[(i)]
    \item $\hat{C}_i(\{s\})=\{s\}$ for some $s\in \{s^1, s^2, \dots , s^t\}\setminus \{\psi[\gamma](i)\}$ such that $C_i(\psi[\gamma](i) \cup \{s\})=\{s\}$, and 
    \item $\hat{C}_i(\{s'\})= \emptyset $ for all $s'\in \{s^1, s^2, \dots , s^t\}\setminus \{s\}$.
\end{enumerate}

By individual rationality and weak non-wastefulness, for any mechanism $\psi$ we have $\psi[\hat{\gamma}](i)=\{s\}$. With $\psi[\gamma](i) \neq \psi[\hat{\gamma}](i)$  and $C_i(\psi[\hat{\gamma}](i)\cup \psi[\gamma](i))= \psi[\hat{\gamma}](i)$, $\psi$ therefore violates incentive compatibility.

\bigskip

\noindent \textbf{Part 2. Acceptable-consistency is necessary.}\\ 
Suppose that for some $C_i$ we have $C_i(\{s^1\})=\{s^1\}$  and $C_i(\{s^2\})=\emptyset$ but $C_i(\{s^1,s^2\})= \{s^2\}$.

Consider an admissions problem $\tilde{\gamma} =\langle I, S, q, C, \tilde{\pi} \rangle$ such that $i \mathrel{\tilde{\pi}}_s i'$ for all $i'\in I\setminus \{i\}$ and $s \in \{ s^1,s^2\}$ and $\emptyset \mathrel{\tilde{\pi}}_s i$ for all $s \in S \setminus  \{ s^1,s^2\}$. Moreover, $\tilde{\pi}$ is such that $|\{i'\in I \setminus\{i\}: i' \mathrel{\tilde{\pi}}_s \emptyset \}|< q_s$ for all $s \in \{ s^1,s^2\}$.

If $\psi[\gamma](i)=\{s\}$ for $s\in S\setminus \{s^1,s^2\}$ we have a violation of individual rationality for the appropriate institution. By weak non-wastefulness we have that for any mechanism $\psi[\gamma](i)\neq \emptyset$
and by individual rationality we have $\psi[\gamma](i)\neq s^2$. Therefore we have $\psi[\gamma](i)=\{s^1\}$.

Consider the following problem $\hat{\gamma}=\langle I, S, q, (\hat{C}_i, C_{-i}), \tilde{\pi} \rangle$  with $\hat{C}_i$ as follows:
\begin{enumerate}[(i)]
    \item $\hat{C}_i(\{s^2\})=\{s^2\}$, and 
    \item $\hat{C}_i(\{s\})=\emptyset$ for all $s\in S \setminus \{s^2\}$
\end{enumerate}

By individual rationality and weak non-wastefulness, for any mechanism $\psi$ we have $\psi[\hat{\gamma}](i)=\{s^2\}$.  With $\psi[\gamma](i) \neq \psi[\hat{\gamma}](i)$  and $C_i(\psi[\hat{\gamma}](i)\cup \psi[\gamma](i))= \psi[\hat{\gamma}](i)$, $\psi$ therefore violates incentive compatibility.
\end{proof}

\subsection*{\Cref{Application_Simultaneous}}
Slightly abusing notation, if we write $\{s\}$ for $s\in S\cup \{\emptyset\}$, then for the case where $s = \emptyset$, $\{\emptyset\}$ should be read as $\emptyset$.

\begin{lem}
\label{pathindependence2}
Consider a path independent choice function $C_i$, then choice from a set is pairwise preferred to institutions in the set, that is, 
$$C_i(S)=\{s\} \implies C_i(\{s,s'\})= \{s\} \quad \text{for all} \quad s' \in S \setminus \{s\}.$$

\end{lem}

\begin{proof}
Suppose $C_i(S)=\{s\}$ and $C_i(\{s,s'\})\neq \{s\}$ for some $s' \in S \setminus \{s\}.$ Then by path independence we have that, $C_i(S)=C_i\big(C_i(\{s,s'\}) \cup (S \setminus \{s,s'\})\big) \neq \{s\}$     since $C_i(\{s,s'\})\neq \{s\}$, a contradiction.

\end{proof}

\noindent We now prove \Cref{Application_Simultaneous}.

\begin{proof}
\textbf{The ``if" part:} If choice functions are path independent, the simultaneous cut-offs mechanism $\psi^{sim}$ yields a stable matching for every admissions problem. 

We start with individual rationality. First, recall that each institution $s$ assigns merit score $m_s(i)$ of $0$ to unacceptable individual $i$. While each institution $s$ announces cut-off $c_s \in \mathbb{N}$. Therefore, no institution admits an unacceptable individual. Second, as shown in \Cref{pathindependence}, each individual $i$ with path independent choice function never chooses an unacceptable alternative; therefore, each individual $i$ is assigned only acceptable institution. That is, for all  for all $\gamma\in \Gamma$, $i\in I$, we have that $\psi^{sim}[\gamma](i)\neq \emptyset$ implies $C_i(\{\psi^{sim}[\gamma](i)\})=\{\psi^{sim}[\gamma](i)\}$.

Moving on to the blocking pairs. First, an individual $i$ can never block with an institution $s$ that never proposed to her during the simultaneous cut-offs algorithm, as in that case either the institution has filled its capacity with individuals that have a higher merit score than $i$ and/or individual $i$ is unacceptable. Second, due to path independence, individual $i$ will never block an acceptable institution with an unacceptable one. Combining both observations, it suffices to consider the sequence of sets of institutions proposing to individual $i$ during the simultaneous cut-offs algorithm. For some admission problem $\gamma\in \Gamma$ and individual $i$ let $(S^1,\dots, S^K)[\gamma](i)$ denote the sequence of sets of institutions proposing to $i$ during the simultaneous cut-offs mechanism. That is, $S^1$ proposes first and so on and so forth until $S^K$, which proposes last. 
  
 If the sequence contains only empty sets, then no acceptable institution is willing to block with $i$.

If each set in the sequence contains (only) unacceptable institutions, then no acceptable institution is willing to block with $i$. If some sets in the sequence contain some acceptable institutions, then due to \Cref{pathindependence} the choice of $i$ from such sets is an acceptable institution. Therefore, under the simultaneous cut-offs algorithm, the first tentatively assigned institution to individual $i$ must be acceptable. Afterward, due to \Cref{pathindependence2}, individual $i$ holds a C-maximal institution at each step of the simultaneous cut-offs algorithm. Therefore, the outcome $\psi^{sim}[\gamma](i)\in (S^1 \cup \dots \cup S^K)$ is such that $C_i(\{\psi^{sim}[\gamma](i),s'\})=\{\psi^{sim}[\gamma](i)\}$ for all $s'\in (S^1 \cup \dots \cup S^K)\setminus \{\psi^{sim}[\gamma](i)\}$.

\bigskip

\noindent \textbf{The ``only if" part:} The simultaneous cut-offs mechanism $\psi^{sim}$ yields a stable matching for every admissions problem only if the choice functions are path independent. The contrapositive is, if some choice functions are not path independent then the simultaneous cut-offs mechanism $\psi^{sim}$ is not stable, that is, there exists at least one admission problem $\gamma\in \Gamma$ for which the outcome  $\psi^{sim}[\gamma]\in \mathcal{M}$ is not stable.

\bigskip

\noindent Consider a violation of path independence $C_i(S'\cup S'')\neq C_i(C_i(S')\cup S'')$ for individual $i$. There are three possibilities:
\begin{enumerate}[(a)]
\item $C_i(S'\cup S'') = \{s^1\}$ and $C_i(C_i(S')\cup S'')) = \{s^2\}$ with $s^1\neq s^2$,
\item $C_i(S'\cup S'') = \emptyset$ and $C_i(C_i(S')\cup S'')) = \{s^2\}$, and
\item $C_i(S'\cup S'') = \{s^1\}$ and $C_i(C_i(S')\cup S'')) = \emptyset$.
\end{enumerate}

\bigskip

\noindent Consider the following admission problems that will lead to the necessary violations:
\begin{enumerate}[(i)]
\item admissions problem $\tilde{\gamma}^1 \in \Gamma$ such that $i \mathrel{\tilde{\pi}^1}_s i'$ for all $i'\in I\setminus \{i\}$ and $s \in S'\cup S''$ and $\emptyset  \mathrel{\tilde{\pi}^1}_s i$ for all $s \in S \setminus ( S'\cup S'')$.

We have $\psi^{sim}[\tilde{\gamma}^1](i)=C_i(S'\cup S'')$ as all institutions in $S'\cup S''$ propose to $i$ in the first round and no other institution proposes.

\item admissions problem $\tilde{\gamma}^2 \in \Gamma$ such that $i \mathrel{\tilde{\pi}^2}_s i'$ for all $i'\in I\setminus \{i\}$ and $s \in C_i(S')\cup S''$ and $\emptyset  \mathrel{\tilde{\pi}^2}_s i$ for all $s \in S \setminus ( C_i(S')\cup S'')$.

We have $\psi^{sim}[\tilde{\gamma}^2](i)=C_i(C_i(S')\cup S''))$ as all institutions in $C_i(S')\cup S''$ propose to $i$ in the first round and no other institution proposes.

\item admissions problem $\tilde{\gamma}^3 \in \Gamma$ such that $i \mathrel{\tilde{\pi}^3}_s i'$ for all $i'\in I\setminus \{i\}$ and $s \in S'$ and $\emptyset  \mathrel{\tilde{\pi}^3}_s i$ for all $s \in S \setminus S'$.

We have $\psi^{sim}[\tilde{\gamma}^3](i)=C_i(S')$ as all institutions in $S'$ propose to $i$ in the first round and no other institution proposes.

\item admissions problem $\tilde{\gamma}^4 \in \Gamma$ such that $i \mathrel{\tilde{\pi}^4}_s i'$ for all $i'\in I\setminus \{i\}$ and $s \in \{s^1, s^2\}$ and $\emptyset  \mathrel{\tilde{\pi}^4}_s i$ for all $s \in S \setminus \{s^1, s^2\}$.

We have $\psi^{sim}[\tilde{\gamma}^4](i)=C_i(\{s^1, s^2\})$ as all institutions in $\{s^1, s^2\}$ propose to $i$ in the first round and no other institution proposes.

\end{enumerate}

\bigskip

\noindent First consider possibility (a), that is, $C_i(S'\cup S'') = \{s^1\}$ and $C_i(C_i(S')\cup S'')) = \{s^2\}$. Then, the following cases can occur.

\begin{itemize}
    \item  \textbf{Case 1. Suppose $C_i(\{s^1\})=\emptyset$.}\\
   $\psi^{sim}[\tilde{\gamma}^1](i)=\{s^1\}$ gives a violation of individual rationality as $s^1$ is unacceptable.

    \item  \textbf{Case 2. Suppose $C_i(\{s^2\})=\emptyset$.}\\
    $\psi^{sim}[\tilde{\gamma}^2](i)=\{s^2\}$ gives a violation of individual rationality as $s^2$ is unacceptable.

    \item  \textbf{Case 3. Suppose $C_i(\{s^1,s^2\})=\emptyset$.}\\
    $\psi^{sim}[\tilde{\gamma}^4](i)=\emptyset$ is blocked by pair $(i,s^1)$ as $s^1\in S'\cup S''$ --- individual $i$ has the highest priority at $s^1$.

    \item  \textbf{Case 4. Suppose $C_i(\{s^1,s^2\})=\{s^2\}$.}\\
    $\psi^{sim}[\tilde{\gamma}^1](i)=\{s^1\}$ is blocked by pair $(i,s^2)$ as $s^2\in S'\cup S''$ --- individual $i$ has the highest priority at $s^2$.

\item  \textbf{Case 5. Suppose $C_i(\{s^1,s^2\})=\{s^1\}$ and $s^1\in C_i(S')\cup  S''$.\\}
$\psi^{sim}[\tilde{\gamma}^2](i)=\{s^2\}$ is blocked by pair $(i,s^1)$ as $s^1\in C_i(S')\cup  S''$ --- individual $i$ has the highest priority at $s^1$. 

\item   \textbf{Case 6. Suppose $C_i(\{s^1,s^2\})=\{s^1\}$ and $s^1\in S'\setminus C_i(S') $ and $s^2=C_i(S')$.\\}
$\psi^{sim}[\tilde{\gamma}^3](i)=\{s^2\}$ is blocked by pair $(i,s^1)$ as $s^1\in S'$ --- individual $i$ has the highest priority at $s^1$. 

\item \textbf{Case 7. Suppose $C_i(\{s^1,s^2\})=\{s^1\}$ and $s^1\in S'\setminus C_i(S')$ and $s^2\in S''$.}

\begin{itemize}
    \item  \textbf{Case 7.1. Suppose $C_i(S')= \{s^3\}$ and $C_i(\{s^1,s^3\})=\{s^1\}$.\\}
$\psi^{sim}[\tilde{\gamma}^3](i)=\{s^3\}$ is blocked by pair $(i,s^1)$ as $s^1\in S'$  --- individual $i$ has the highest priority at $s^1$. 

    \item  \textbf{Case 7.2. Suppose $C_i(S')= \{s^3\}$ and $C_i(\{s^1,s^3\})=\{s^3\}$.\\}
 $\psi^{sim}[\tilde{\gamma}^1](i)=\{s^1\}$ is blocked by pair $(i,s^3)$ as $s^3\in S'\cup S''$ --- individual $i$ has the highest priority at $s^3$. 

    \item  \textbf{Case 7.3. Suppose $C_i(S')= \{s^3\}$ and $C_i(\{s^1,s^3\})=\emptyset$.\\}
 Then, the problem is similar to Case 3.

   \item  \textbf{Case 7.4. Suppose $C_i(S') = \emptyset$.\\}
$\psi^{sim}[\tilde{\gamma}^3](i)=\emptyset$ is blocked by pair $(i,s^1)$ as $s^1\in S'$  ---  individual $i$ has the highest priority at $s^1$.

\end{itemize}
\end{itemize}

\bigskip
\noindent Next, consider possibility (b), that is, $C_i(S'\cup S'') = \emptyset$ and $C_i(C_i(S')\cup S'')) = \{s^2\}$. Then, the following cases can occur.

\begin{itemize}
    \item  \textbf{Case 1. Suppose $C_i(\{s^2\})=\emptyset$.}\\
    $\psi^{sim}[\tilde{\gamma}^2](i)=\{s^2\}$ gives a violation of individual rationality as $s^2$ is unacceptable.
    
    \item  \textbf{Case 2. Suppose $C_i(\{s^2\})=\{s^2\}$.}\\
    $\psi^{sim}[\tilde{\gamma}^1](i)=\emptyset$ is blocked by pair $(i,s^2)$ as $s^2\in S'\cup S''$ --- individual $i$ has the highest priority at $s^2$.

\end{itemize}

\bigskip
\noindent Lastly, consider possibility (c), that is, $C_i(S'\cup S'') = \{s^1\}$ and $C_i(C_i(S')\cup S'')) = \emptyset$. Then, the following cases can occur.

\begin{itemize}
    \item  \textbf{Case 1. Suppose $C_i(\{s^1\})=\emptyset$.}\\
  $\psi^{sim}[\tilde{\gamma}^1](i)=\{s^1\}$ gives a violation of individual rationality as $s^1$ is unacceptable.

 \item  \textbf{Case 2. Suppose $C_i(\{s^1\})=\{s^1\}$.}  and $s^1\in C_i(S')\cup  S''$.\\
    $\psi^{sim}[\tilde{\gamma}^2](i)=\emptyset$ is blocked by pair $(i,s^1)$ as $s^1\in C_i(S')\cup  S''$ --- individual $i$ has the highest priority at $s^1$. 

\item   \textbf{Case 3. Suppose $C_i(\{s^1\})=\{s^1\}$. and $s^1\in S'\setminus C_i(S') $.}
\begin{itemize}
    \item  \textbf{Case 3.1. Suppose $C_i(S')= \{s^3\}$ and $C_i(\{s^1,s^3\})=\{s^1\}$.\\}
$\psi^{sim}[\tilde{\gamma}^3](i)=\{s^3\}$ is blocked by pair $(i,s^1)$ as $s^1\in S'$  --- individual $i$ has the highest priority at $s^1$. 

    \item  \textbf{Case 3.2. Suppose $C_i(S')= \{s^3\}$ and $C_i(\{s^1,s^3\})=\{s^3\}$.\\}
 $\psi^{sim}[\tilde{\gamma}^1](i)=\{s^1\}$ is blocked by pair $(i,s^3)$ as $s^3\in S'\cup S''$ --- individual $i$ has the highest priority at $s^3$. 

    \item  \textbf{Case 3.3. Suppose $C_i(S')= \{s^3\}$ and $C_i(\{s^1,s^3\})=\emptyset$.\\}
Then, the problem is similar to Case 3 under possibility (a).

   \item  \textbf{Case 3.4. Suppose $C_i(S') = \emptyset$.\\}
$\psi^{sim}[\tilde{\gamma}^3](i)=\emptyset$ is blocked by pair $(i,s^1)$ as $s^1\in S'$  ---  individual $i$ has the highest priority at $s^1$. 
\end{itemize}

\end{itemize}
\end{proof}

\subsection*{\Cref{Application_Sequential}}

Slightly abusing notation, if we write $\{s\}$ for $s\in S\cup \{\emptyset\}$, then for the case where $s = \emptyset$, $\{\emptyset\}$ should be read as $\emptyset$.

\begin{lem}
\label{acyclicity2}
Consider an acyclic and acceptable-consistent choice function $C_i$ and a finite sequence of institutions $s^1, \dots, s^K$ with $K\geq 2$ that propose to individual $i$ (one at a time) under the sequential cut-offs algorithm. That is, $s^1$ proposes first and so on and so forth until $s^K$, which proposes last. If the individual's last tentative assignment is $s^*$, then $C_i(\{s^k\} \cup \{s^*\})\neq \{s^k\}$ for any $s^k \in\{s^1, \dots, s^K\} \setminus \{s^*\}$.

\end{lem}

\begin{proof}
If all institutions are unacceptable, then under sequential cut-offs algorithm $s^*=\emptyset$.
Given our abuse of notation, we have  $C_i(\{s^k\} \cup \{s^*\})= C_i(\{s^k\} \cup \{\emptyset\})=C_i(\{s^k\} \cup \emptyset) = C_i(\{s^k\})$.
As all institutions are unacceptable, $C_i(\{s^k\})= \emptyset \neq \{s^k\}$ for all $s^k \in\{s^1, \dots, s^K\}$.

If some institutions are acceptable, then under the sequential cut-offs algorithm $s^* \neq \emptyset$ and $C_i(\{s^*\})= \{s^*\}$, that is $s^*$ is an acceptable alternative. This is because the first alternative tentatively assigned to any individual $i$ must be acceptable. Afterward, under the sequential cut-offs algorithm, individual $i$'s assignment is updated only if another alternative is chosen over $i$'s tentative assignment. Due to acceptable-consistency, only an acceptable alternative can be chosen over $i$'s tentative assignment.

It remains to be shown that for the final tentative assignment $s^*$, we have $C_i(\{s^k\} \cup \{s^*\})\neq \{s^k\}$ for any $s^k \in\{s^1, \dots, s^K\} \setminus \{s^*\}$.
Let $s^{(j)}$ denote the tentative assignment of individual $i$ when it receives proposal from institution $s^j$. Therefore, when institution $s^j$ proposes individual $i$'s choice menu is $\{s^j\} \cup \{s^{(j)} \}$. 

Suppose there exists $s^k \in \{s^1, \dots s^K\} \setminus \{s^*\}$ such that $C_i(\{s^k,s^*\})= \{s^k\}$. By acceptable-consistency, $s^k$ must be an acceptable alternative.
For $K=2$ we have an immediate contradiction with $s^*$ being the last tentative assignment. For $K\geq 3$ there are two possibilities to consider.
\begin{enumerate}[(i)]
    \item Suppose $C_i(\{s^k\} \cup \{s^{(k)} \})=\{s^k\}$. Since $s^k$ is not the last tentative assignment there must exist an acceptable alternative $s^{k'}\in\{s^{k+1}, \dots, s^K\}$ such that $C_i(\{s^k\}\cup \{ s^{k'}\}) = \{s^{k'}\}$. If $s^{k'}=s^*$ we have a contradiction. Otherwise, if $s^{k'}$ is not the last tentative assignment there must exists an acceptable alternative $s^{k''}\in\{s^{k'+1}, \dots, s^K\}$ such that $C_i(\{s^{k'}\}\cup\{ s^{k''}\}) = \{s^{k''}\}$. By acyclicity, $C_i(\{s^k\}\cup \{ s^{k'}\}) = \{s^{k'}\}$ and $C_i(\{s^{k'}\}\cup \{s^{k''}\}) = \{s^{k''}\}$ imply  that $C_i(\{s^{k}\} \cup \{s^{k''}\}) = \{s^{k''}\}$.  If $s^{k''}=s^*$ we have a contradiction. 
    Otherwise,  $s^{k''}$ is not the last tentative assignment. We can finitely repeat the same steps until we arrive at the last tentative assignment, which due to acyclicity, will lead to a contradiction
    with $C_i(\{s^k,s^*\})= \{s^k\}$ as $C_i(\{s^{k}\} \cup \{s^{*}\}) = \{s^{*}\}$
    .
    
    \item Suppose $C_i(\{s^k\} \cup \{s^{(k)} \}) \neq \{s^{k}\}$. If $s^{(k)}=s^*$ we have a contradiction. Otherwise, if $s^{(k)}$ is not the last tentative assignment, there must exist an acceptable alternative $s^{k'}\in\{s^{k+1}, \dots, s^K\}$ such that $C_i(\{s^{(k)}\} \cup \{s^{k'}\}) = \{s^{k'}\}$. If $s^{k'}=s^*$, then $C_i(\{s^k\} \cup \{s^*\})= \{s^k\}$ and $C_i(\{s^{(k)}\} \cup \{s^{k'}\}) = \{s^{k'}\}$ together with acyclicity imply that $C_i(\{s^k\} \cup \{s^{(k)} \}) = \{s^{k}\}$, therefore we have a contradiction. Otherwise, if $s^{k'}$ is not the last tentative assignment there must exists an acceptable alternative $s^{k''}\in\{s^{k'+1}, \dots s^K\}$ such that $C_i(\{s^{k'}\} \cup \{s^{k''}\}) = \{s^{k''}\}$. By acyclicity, $C_i(\{s^{(k)}\} \cup \{s^{k'}\}) = \{s^{k'}\}$ and $C_i(\{s^{k'}\} \cup \{s^{k''}\}) = \{s^{k''}\}$ imply that $C_i(\{s^{(k)}\} \cup \{s^{k''}\}) = \{s^{k''}\}$.  If $s^{k''}=s^*$, then by acyclicity $C_i(\{s^k\} \cup  \{s^*\})= \{s^k\}$ and $C_i(\{s^{(k)}\} \cup \{s^{k''}\}) = \{s^{k''}\}$ imply that $C_i(\{s^k\} \cup \{s^{(k)} \}) = \{s^{k}\}$, therefore we have a contradiction. Otherwise, $s^{k''}$ is not the last tentative assignment. We can finitely repeat the same steps until we arrive at the last tentative assignment, which due to acyclicity will lead to a contradiction with $C_i(\{s^k\} \cup \{s^{(k)} \}) \neq \{s^{k}\}$ as  $C_i(\{s^k\} \cup \{s^{(k)} \}) = \{s^{k}\}$.
\end{enumerate}

\end{proof}

\noindent We now prove \Cref{Application_Sequential}.

\begin{proof}
\textbf{The ``if" part:}   If choice functions are acyclic and acceptable-consistent, the sequential cut-offs mechanism $\psi^{seq}$ yields a stable matching for every admissions problem.

We start with individual rationality. In the sequential cut-offs algorithm, each institution $s$ announces cut-offs $c_s \in \mathbb{N}$, while any unacceptable individual $i\in I$ at $s$ has a merit score of $m_s(i)=0$ and therefore can never be assigned to $s$. Meanwhile, as shown in \Cref{acyclicity2}, each individual's last tentative assignment is either an acceptable institution or it remains unassigned.

Moving on to the blocking pairs. First, an individual $i$ can never block with an acceptable institution $s\in S$ that never proposed to her during the sequential cut-offs algorithm, as in that case either the institution has filled its capacity with individuals that have a higher merit score than $i$ and/or individual $i$ is unacceptable. Second, due to acceptable-consistency, individual $i$ will never block its assigned institution with an unacceptable institution.
Combining these observations, it suffices to consider the sequence of institutions proposing to individual $i$ during the sequential cut-offs algorithm. 

If no institution or only one institution proposes to some $i\in I$ then there can be no blocking pair involving individual $i$ and an institution. Otherwise, by \Cref{acyclicity2}, the outcome $\psi^{seq}[\gamma](i)\in \{s^1,\dots, s^K\}$ is such that $C_i(\{\psi^{seq}[\gamma](i),s'\})\neq \{s'\}$ for all $s'\in \{s^1,\dots, s^K\}\setminus \{\psi^{seq}[\gamma](i)\}$.

\bigskip

\noindent \textbf{The ``only if" part:}  The sequential cut-offs mechanism $\psi^{seq}$ yields a stable matching for every admissions problem only if the choice functions are acyclic and acceptable-consistent. The contrapositive is, if some choice functions are not acyclic and acceptable-consistent then the sequential cut-offs mechanism $\psi^{seq}$ is not stable, that is, there exists at least one admission problem $\gamma\in \Gamma$ for which the outcome  $\psi^{seq}[\gamma]\in \mathcal{M}$ is not stable.

\bigskip

\noindent \textbf{Part 1. Acylicity is necessary.}\\
\textbf{Case 1:} Suppose that for some $C_i$ we have a cycle for $t\geq 3$ and distinct (acceptable) institutions $s^1, s^2, \dots , s^t$ in $S$ such that  $C_i(\{s^1\} \cup \{s^2\}) = \{s^2\}, \dots, C_i(\{s^{t-1}\} \cup \{s^{t}\}) = \{s^{t}\}, \text{ and } C_i(\{s^{1}\} \cup \{s^{t}\}) = \{s^{1}\}$.
Consider an admission problem $\tilde{\gamma} \in \Gamma$ such that $i \mathrel{\tilde{\pi}}_s i'$ for all $i'\in I\setminus \{i\}$ and $s \in \{ s^1, s^2, \dots , s^t\}$ and $\emptyset \mathrel{\tilde{\pi}}_s i$ for all $s \in S \setminus \{ s^1, s^2, \dots , s^t\}$.
  Given the definition of the sequential cut-offs mechanism, we have that $\psi^{seq}[\tilde{\gamma}](i) \in \{ s^1, s^2, \dots , s^t\}$, regardless of outcome we have a blocking pair proving the claim. 
If $\psi^{seq}[\tilde{\gamma}](i)=\emptyset$ the same holds, as all institutions in  $\{ s^1, s^2, \dots , s^t\}$ are acceptable, leading to a blocking pair.

\bigskip

\noindent \textbf{Case 2:} Suppose that for some $C_i$ we have a weak cycle for $t\geq 3$ and distinct (acceptable) institutions $s^1, s^2, \dots, s^t$ in $S$ such that  $C_i(\{s^1\} \cup \{s^2\}) = \{s^2\}, \dots, C_i(\{s^{t-1}\} \cup \{ s^{t}\}) = \{s^{t}\}, \text{ and } C_i(\{s^{1}\} \cup \{s^{t}\}) = \emptyset $.
Consider an admission problem $\tilde{\gamma} \in \Gamma$ such that $i \mathrel{\tilde{\pi}}_s i'$ for all $i'\in I\setminus \{i\}$ and $s \in \{ s^1, s^2, \dots , s^t\}$ and $\emptyset \mathrel{\tilde{\pi}}_s i$ for all $s \in S \setminus \{ s^1, s^2, \dots , s^t\}$.
  In the sequential cut-offs mechanism, $s^1, s^2, \dots,  s^t$ will propose to $i$ and no other institution will propose to $i$. 
Consider the proposal sequence $s^1, s^t, s^2, \dots, s^{t-1}$.
That is, $s^1$ proposes before $s^t$, $s^t$ proposes before $s^2$, $s^2$ proposes before $s^3$, and so on until $s^{t-1}$. 
We get that the outcome of the sequential cut-offs mechanism is $\psi^{seq}[\tilde{\gamma}](i)=\{s^{t-1}\}$. But in this case there is a blocking pair as $C_i(\{s^{t-1}\} \cup \{s^t\})= \{s^t\}$.

\bigskip

\noindent \textbf{Part 2. Acceptable-consistency is necessary.}\\ 
Suppose that for some $C_i$ we have $C_i(\{s^1\})=\{s^1\}$  and $C_i(\{s^2\})=\emptyset$ but $C_i(\{s^1\} \cup \{s^2\})= \{s^2\}$.
Consider an admission problem $\tilde{\gamma} \in \Gamma$ such that $i \mathrel{\tilde{\pi}}_s i'$ for all $i'\in I\setminus \{i\}$ and $s \in \{ s^1,s^2\}$ and $\emptyset \mathrel{\tilde{\pi}}_s i$ for all $s \in S \setminus  \{ s^1,s^2\}$.
In the sequential cut-offs mechanism both $s^1$ and $s^2$ will propose to $i$. 
If $s^2$ proposes first, we have that $\psi^{seq}[\tilde{\gamma}](i)=\{s^1\}$. But, there is a blocking pair $(i,s^2)$ as $C_i(\{s^1\} \cup \{s^2\})= \{s^2\}$ and $i$ has the highest priority at $s^2$.
Otherwise, if $s^1$ proposes first we have that $\psi^{seq}[\tilde{\gamma}](i)=\{s^2\}$. But, there we have a violation of individual rationality as as $C_i(\{s^2\})= \emptyset$.
\end{proof}

\subsection*{\Cref{groupstability}}
Recall that, from an institutional viewpoint there are no complementarities between individuals, so the priority order $\pi_s$ and capacity $q_s$ of an institution $s$ translate into a (partial order) preference over sets of individuals in a straightforward way. Formally, let $\succsim_s$ denote the preferences of institution $s$ over $2^I$, and $\succ_s$ denote strict preferences derived from it. We assume that college preferences are responsive. Formally, $\succsim_s$ is \textbf{responsive} (\cite{roth1985college}) if, \begin{enumerate}[(i)]
    \item for any $I' \subset I$ with $|I'|<q_s$ and any $i \in I \setminus I'$, 
$$ (I' \cup \{i\}) \mathrel{\succ_s} I' \iff i \mathrel{\pi_s} \emptyset,$$   

\item for any $I' \subset I$ with $|I'|<q_s$ and any $i,i' \in I \setminus I'$, 
$$ (I' \cup \{i\}) \mathrel{\succ_s} (I' \cup \{i'\}) \iff i \mathrel{\pi_s} i'.$$   
\end{enumerate} 

We say matching $\mu$ is \textbf{blocked by a coalition $T$} of individuals and institutions, if there exists another matching $\nu$ and coalition $T$, such that for all $i \in T$ and $s\in T$,
\begin{enumerate}[(i)]
    \item $\nu(i) \in T$,
    \item $C_i(\nu(i) \cup\mu(i) )=\nu(i)$ for all $i \in T$,
    \item $\nu(s) \mathrel{\succ_s} \mu(s)$ for all $s \in T$, and
    \item if $j \in \nu(s)$, then $j \in T \cup \mu(s)$.
\end{enumerate} 
The first condition states that every individual in $T$ who is matched by $\nu$ is matched to some institution in $T$. The second condition states that every individual in $T$ chooses its assignment under $\nu$ over her assignment under $\mu$. The third condition states that every institution in $T$ strictly prefers its set of individuals under $\nu$ to that under $\mu$. The last condition states that any new individual matched to an institution in the coalition must be a member of $T$.

A \textbf{group stable} matching is one that is not blocked by any coalition. Pairwise stability is equivalent to group stability in the standard setup of many-to-one matching markets (\cite{roth_sotomayor_1990}).  The same result  holds in our setup. 

\begin{prop}\label{groupstability}
A matching is group stable if and only if it is stable.
\end{prop}
\begin{proof}
Suppose $\mu$ is not stable due to an unacceptable individual (institution) assigned to an institution (individual), or a blocking pair. Then it is not group stable because it is blocked by the coalition consisting of the  individual (institution), or the blocking pair respectively.

In the other direction, if $\mu$ is blocked by coalition $T$ and matching $\nu$. 
We show that we can always construct a blocking pair or a violation of individual rationality from $T$.

Suppose no institution is part of $T$, then there exists an individual $i\in T$ such that $\nu(i)=\emptyset$. By definition of the blocking coalition we have $C_i(\nu(i) \cup \mu(i) )=\emptyset$ with $\mu(i)\in S$. Therefore, we have found a violation of individual rationality and thus pairwise stability. 

Suppose no individual is part of $T$, then there exists an institution $s\in T$ such that $\nu(s)\subset \mu(s)$. By definition of the blocking coalition we have $\nu(s) \succ_s \mu(i)$ and by responsiveness some $i\in \mu(i)\setminus \nu(i)$ with $\emptyset \pi_s i$. Therefore, we have found a violation of individual rationality (on the institution side) and thus pairwise stability. 

Finally, the coalition contains both individuals and institutions. 
Then pick institution $s\in T$ with $\nu(s)\succ_s \mu(s)$. 
Suppose there exists at least an individual $i \in \nu(s) \setminus \mu(s)$. Then we either have (i) $i \succ_s j$ for some $j \in \mu(s) \setminus \nu(s)$ or (ii) $i \succ_s \emptyset$ and $|\mu(s)|$. By the definition of a blocking coalition we have  $C_i(\{s\} \cup \mu(i) )=\{s\}$, and therefore $i$ and $s$ form a blocking pair.
Otherwise, if $\nu(s)\subset \mu(s)$ we again can construct a violation of individual rationality (on the institution side).
\end{proof}

\end{document}